\RequirePackage{fix-cm}
\documentclass[smallextended]{svjour3}    
\smartqed  
\usepackage{graphicx}
\usepackage{float}
\usepackage{mathptmx}      
\usepackage{amsmath}
\usepackage{siunitx}
\usepackage[title]{appendix}
\newcommand\ddfrac[2]{\frac{\displaystyle #1}{\displaystyle #2}}

\journalname{Experimental Astronomy}

\begin{document}

\title{SARAS 2: A Spectral Radiometer for probing Cosmic Dawn and the Epoch of Reionization through detection of the global 21 cm signal}

\titlerunning{SARAS 2: radiometer probing CD and EoR} 

\author{Saurabh Singh \and Ravi Subrahmanyan \and N. Udaya Shankar \and Mayuri Sathyanarayana Rao \and B.S. Girish\and A. Raghunathan \and R. Somashekar \and K.S. Srivani}

\institute{Saurabh Singh \and Ravi Subrahmanyan \and N. Udaya Shankar \and Mayuri Sathyanarayana Rao \and B.S. Girish\and A. Raghunathan \and R. Somashekar \and K.S. Srivani \at
Raman Research Institute, C V Raman Avenue, Sadashivanagar, Bangalore 560080, India \\ \\
Saurabh Singh \at
\email{saurabhs@rri.res.in} \\ \\
Saurabh Singh \at
Joint Astronomy Programme, Indian Institute of Science, Bangalore 560012, India
}
\date{Received: date / Accepted: date}
\maketitle
\begin{abstract}

The global 21 cm signal from Cosmic Dawn (CD) and the Epoch of Reionization (EoR), at redshifts $z \sim 6-30$, probes the nature of first sources of radiation as well as physics of the Inter-Galactic Medium (IGM). Given that the signal is predicted to be extremely weak, of wide fractional bandwidth, and lies in a frequency range that is dominated by Galactic and Extragalactic foregrounds as well as Radio Frequency Interference, detection of the signal is a daunting task. Critical to the experiment is the manner in which the sky signal is represented through the instrument. It is of utmost importance to design a system whose spectral bandpass and additive spurious can be well calibrated and any calibration residual does not mimic the signal. SARAS is an ongoing experiment that aims to detect the global 21 cm signal. Here we present the design philosophy of the SARAS 2 system and discuss its performance and limitations based on laboratory and field measurements.   Laboratory tests with the antenna replaced with a variety of terminations, including a network model for the antenna impedance, show that the gain calibration and modeling of internal additives leave no residuals with Fourier amplitudes exceeding 2~mK, or residual Gaussians of 25 MHz width with amplitudes exceeding 2~mK.  Thus, even accounting for reflection and radiation efficiency losses in the antenna, the SARAS~2 system is capable of detection of complex 21-cm profiles at the level predicted by currently favoured models for thermal baryon evolution.

\keywords{Astronomical instrumentation \and Methods: observational \and Cosmic background radiation \and Cosmology: observations \and Dark ages \and Reionization \and First stars}

\end{abstract}

\section{Introduction}
\label{intro}

Cosmic Dawn (CD) and the following Epoch of Reionization (EoR) mark important turning points in the thermal and ionization state of baryons. Following cosmological recombination, the predominantly neutral gas in the Universe is completely ionized by $z\sim 6$ \cite{2006ARA&A..44..415F,2015MNRAS.447..499M,1538-4357-617-1-L5,1538-3881-122-6-2850}. However, the nature of the first sources of radiation that transformed the thermal and ionization state of the gaseous baryons in the Universe, their properties, as well as a precise timeline of various phases of heating and reionization, are poorly constrained. Detection of 21 cm radiation from neutral Hydrogen during this epoch directly captures the underlying astrophysics and hence is a potential tool to address these long-standing issues \cite{doi:10.1146/annurev-astro-081309-130936,0004-637X-782-2-66,1538-4357-624-2-L65,doi:10.1111/j.1365-2966.2007.11519.x}.

The 21 cm signal from CD and EoR can be studied through its brightness temperature fluctuations over spatial and spectral domains, which includes a sky-averaged or global signal varying across frequency. The latter represents a mean departure of the 21 cm brightness temperature from the cosmic microwave background (CMB) and hence sets a reference level for fluctuation measurements. In addition, the global signal also encapsulates astrophysical information about the nature of sources as well as the sequence of events during CD and EoR \cite{PhysRevD.82.023006,2013ApJ...777..118M,2015ApJ...813...11M,doi:10.1111/j.1365-2966.2005.09485.x,doi:10.1111/j.1365-2966.2005.09485.x}. Since it is an all-sky/global signal, high spatial resolution is not needed and a well-calibrated and efficient single-element radiometer can achieve the required sensitivity in a few minutes \cite{2017ApJ...840...33S}. However, the detection of the signal is challenging owing to multiple reasons: the signal is predicted to be extremely weak, with maximum amplitude less than $300~\rm mK$ in brightness temperature, smoothly varying over a wide frequency range from about $200~\rm MHz$ all the way to below $40~\rm MHz$ \cite{2016arXiv160902312C} and buried in Galactic and Extragalactic foregrounds of $100-10,000~\rm K$ \cite{1999A&A...345..380S,2017ApJ...840...33S}. 

The major challenge is with the design of the radiometer because first the sensor may couple structure in the sky into spectral modes, and second the sensor and receiver modify the shape of the incident sky signal by a frequency response that manifests as both multiplicative gain and additive components. Hence, if not modeled adequately, the system response can confuse the detection of the signal through its own spurious and residual signatures. Various design and analysis strategies have been evoked to deal with a variety of system architectures \cite{2017ApJ...835...49M,2015PASA...32....4S,2014ApJ...782L...9V,2017arXiv170909313P} and modeling of their response.

SARAS aims at detecting the global 21 cm signal from CD and subsequent EoR in the frequency range of $40-200~\rm MHz$; this paper describes the development, architecture and performance tests of the SARAS 2 radiometer. It is a single element spectral radiometer that employs an antenna with a frequency independent beam and a noise source for calibrating the system. It provides a differential measurement between the antenna temperature and a reference load. A splitter is used to divide the signals from the antenna and noise source into two paths and the final spectrum is obtained by cross correlating the signals in these two paths.  Further, this measurement is phase switched to cancel spurious additives in the system. The signal in the two paths are transmitted to a signal conditioning unit, placed 100 m from the antenna, over optical fibers thereby providing optical isolation. All sub-systems are designed with the aim of making the different contributions -- additive and multiplicative -- to have smoothly varying functional forms that might not confuse with plausible forms of the global EoR signal. 

In Section 2, we provide a brief overview of the SARAS 2 system. In Sections $3-5$, we discuss different sub-systems of the SARAS 2 system --- the antenna, analog signal processor and the digital signal processor --- and the underlying design considerations that led to their final adopted configuration. We also discuss the advantages and limitations of the present architecture based on laboratory and field measurements. We describe the software developed to process the data in Section~6, which includes calibration and flagging,  and discuss the rationale behind the algorithms used, most of which were custom developed for SARAS 2. In Section~7 we evaluate system performance using a set of terminations replacing the antenna. In Section~8, we compare the architecture of the SARAS 2 system with that of other ongoing experiments to detect the global 21 cm from CD and EoR;  a summary is in Section~9.

\section{Overview}
\label{sec:1}

SARAS (Shaped Antenna measurement of the background RAdio Spectrum) is a spectral radiometer that aims to detect redshifted 21 cm radiation from CD and EoR. The first version of the instrument, SARAS 1, provided an improved calibration for the $150~\rm MHz$ all-sky map of Landecker \& Wielebinski \cite{0004-637X-801-2-138}. It consisted of a fat-dipole antenna above ferrite-tile absorbers, an analog receiver located at ground level just beneath the antenna, and a digital spectrometer unit 100-m away that provided sky spectra with 1024 channels over the band $87.5 - 175~\rm MHz$ \cite{2013ExA....36..319P}. 

SARAS 2 described herein operates over the frequency band 40--200~MHz.  Observations to date with SARAS 2 have been able to rule out reionization scenarios where the reionization is rapid and the first X-ray sources have very poor heating efficiency \cite{2017ApJ...845L..12S}.  

The first sub-system in the radiometer is a spherical monopole antenna---consisting of a spherical element above a disc---that acts as the sensor of the electromagnetic field.  The total efficiency of the antenna was relatively poorer at low frequencies; therefore the results to date have used data over the band $110 - 200~\rm MHz$ and this paper presents test of system performance over this restricted band.  

\begin{figure}[htbp]
\begin{center}
\includegraphics[scale=0.8]{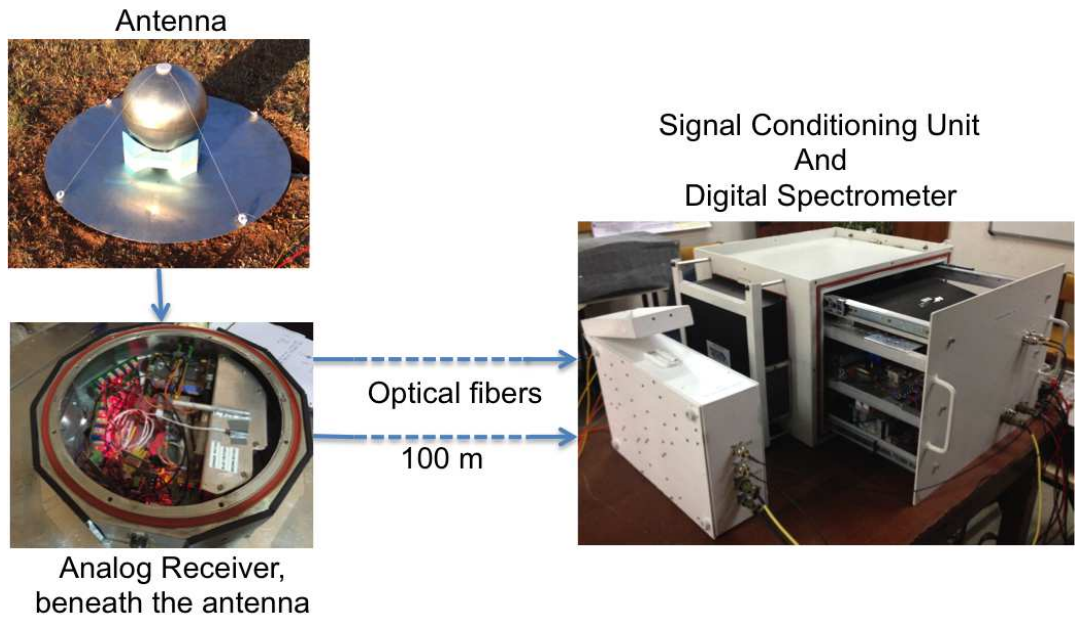}
\caption{SARAS 2 system schematic.}
\label{fig:0}    
\end{center}   
\end{figure}

Beneath the metallic disc of the antenna lies the receiver electronics that splits and amplifies the signals from the antenna, reference and calibration noise source, generates linear combination of these signals and phase switches them before transmitting over optical fibers. 100~m away from the antenna is a signal processing unit that re-converts the optical signals back to radio frequency (RF) and filters out frequencies outside the band of interest. Finally, the signals enter a digital spectrometer that digitizes, resolves the data into narrow spectral channels, and cross correlates them to produce the sky spectra. The schematic of the system is shown in Fig.~\ref{fig:0}. The entire system runs on batteries and can be deployed at remote locations.

Future developments are aimed at improving performance to widen the useful band. 

\section{Antenna}
\label{sec:1a}

\subsection{General considerations for EoR experiments }
\label{sec:gen_ant}

The antenna is one of the critical sub-systems of the entire radiometer. Various antenna properties that affect the data, e.g. the beam pattern, the reflection, radiation and total efficiencies, all vary across the band and require considerable effort and care to measure to the accuracies required to model their effects on the data. Thus it is crucial to pay close attention to the design of the antenna and ensure that its characteristics do not limit the detection of the signal. We will discuss the key antenna properties in the following subsections; in particular how they affect the global EoR measurement.

\subsubsection{Antenna Beam Power Pattern} 
\label{sec:beam_phil}
We denote the sky brightness distribution, weighted by the antenna beam pattern, by $T_W(\nu,t)$; this is a function of frequency $\nu$ and for a radiometer pointed towards fixed azimuth and elevation, the spectrum varies with time $t$ as the sky drifts overhead. It may be written as:
\begin{equation}
T_W(\nu, t)= \ddfrac{\int_{0}^{2\pi} \int_{0}^{\pi} T_B (\theta, \phi, \nu, t) G(\theta, \phi, \nu) \textrm{sin}\theta d\theta d\phi  }{\int_{0}^{2\pi} \int_{0}^{\pi} G(\theta, \phi, \nu) \textrm{sin}\theta d\theta d\phi}.
\label{eq:1}
\end{equation}
$G$ is the antenna beam power pattern over azimuth, $\phi$ and elevation $\theta$, and may be a function of frequency $\nu$.  $T_B$ is the brightness temperature of the sky towards any azimuth and elevation, which varies over time as the sky drifts. The integral is over $4\pi$ steradian accounting for any beam spillover to the ground.

The dominant component of $T_B$ is the Galactic and extra-galactic emission, which we refer to as foregrounds. It arises through various radiative processes and at the frequencies of interest here  is dominated by the synchrotron mechanism. Its absolute contribution is about $3-5$ orders of magnitude larger than the predicted 21 cm signal: while the 21 cm signal is expected to be up to a few hundred mK, the foreground can range from a few hundreds to thousands of Kelvin over the band \cite{1999A&A...345..380S}. 

The foreground in the CD and EoR band has been shown to be a \textit{maximally smooth} function \cite{2015ApJ...810....3S}, which implies that the foreground spectrum can be represented by polynomials that do not have zero crossings in second and higher order derivatives. Any reference to smoothness in this paper assumes this definition. The foregrounds may be fit to the accuracy needed for 21-cm CD/EoR detection using such polynomials, thus leaving more complex components, including a significant part of more complex EoR signals, as residuals \cite{2017ApJ...840...33S}.  

It may be noted here that the 21-cm signal is also expected to have a smooth component which is inseparable from the foreground. Thus when a maximally smooth function is used to model and subtract the foreground, a part of the 21-cm signal is also inevitably erased. Further, since the total efficiency of the antenna may also result in that only a fraction of the sky signal couples into the receiver, we expect an additional loss in the signal. Thus, we have chosen to aim to design a system in which any additive spurious remains below about a $\rm mK$, allowing for substantial signal loss due to these causes.

If the antenna beam pattern $G$ is frequency dependent, then spatial structure in the foreground would couple into the spectral domain and result in a non-smooth spectral response to structure in the continuum sky emission, which can be difficult to model to the accuracy needed for 21 cm signal detection. Thus, it is ideal to have an antenna beam pattern that is independent of frequency; in other words, the beam should be achromatic. However, if the beam is a single lobe, without sidelobes, and whose shape only varies smoothly with frequency, the resulting $T_W$ might still be modeled as a maximally smooth function. 

\subsubsection{Antenna Radiation and Reflection Efficiency}
\label{sec:gen_ref}

Radiation Efficiency, denoted by $\eta_r (\nu)$, determines the fraction of beam-weighted sky power, $T_W$ (Eq.~\ref{eq:1}), that couples into the antenna. Depending on the antenna design, $\eta_r (\nu)$ can vary with frequency. The power obtained after being modified by radiation efficiency, $T_R$, is given by
\begin{equation}
T_R(\nu) = \eta_r(\nu)T_W(\nu).
\end{equation} 

In addition, antennas have impedances that vary with frequency and are differently matched to the connecting transmission line across frequency. This is quantified as reflection coefficient. The voltage reflection coefficient of the antenna, $\Gamma_{c}$, determines how much of $T_R(\nu)$ couples to the system \cite{2012RaSc...47.0K06R}.  The power $T_A$ that propagates along the transmission line connected to the antenna, which we refer to as the antenna temperature, is given by
\begin{equation}
T_A (\nu) = (1-|\Gamma_c (\nu)|^2)T_R(\nu),
\label{eq:2}
\end{equation}
where $T_R$ is the power available at the antenna terminal.  We term the coupling factor, $(1-|\Gamma_c|^2)$, as the reflection efficiency $\eta_c$. Any spectral signature present in $\Gamma_c$ is clearly imprinted on the sky signal through the reflection efficiency.  Therefore, the power or antenna temperature measured by the system in response to the sky brightness is:
\begin{equation}
T_A(\nu) = \eta_r(\nu)(1-|\Gamma_c (\nu)|^2)T_W(\nu).
\end{equation} 
The product $\eta_r(\nu)(1-|\Gamma_c (\nu)|^2)$ is termed as total efficiency $\eta_t$ and hence
\begin{equation}
T_A(\nu) = \eta_t(\nu)T_W(\nu).
\label{eq:etanu}
\end{equation} 

We require $\eta_t$ to be spectrally maximally smooth in order to avoid any complex distortion arising from the multiplicative transfer function represented by the antenna. Further, as discussed below in Sec.~\ref{sec:imp_mis}, $\Gamma_c$ needs to also be spectrally maximally smooth to avoid additive spectral shapes arising from internal systematics.

\subsubsection{Resistive loss}
\label{sec:resis}

Antennas, like dipoles, are balanced sensors and often need to be connected to unbalanced transmission lines such as coaxial lines.
Most such antenna designs use what is called a balun, or balanced to unbalanced transformer, to provide better match between the antenna impedance and that of the connecting transmission line and thus improve reflection efficiency.  The balun also avoids radiation leakage and hence frequency-dependent beam distortions that may arise from unbalanced currents in the connecting cable.

The presence of any such balun almost always results in significant resistive losses that may be complex functions of frequency, particularly over the wide bandwidths needed for CD and EoR detection, and their multiplicative and additive effects on the signal cannot be characterized easily to the required accuracy. Similarly, any loading of antennas to adjust its resonant frequency also leads to resistive losses.  All of these result in additive or multiplicative terms in Eq.~\ref{eq:etanu} depending upon the origin of the resistive loss \cite[Chapter 2]{Balanis:2005:ATA:1208379}. Thus it is best to avoid antenna designs that might have significant resistive losses and also a balun.

\subsection{Evolution to SARAS 2 antenna}

Given the considerations in Sec.~\ref{sec:gen_ant}, we now discuss a variety of classes of antennas that may be suitable for wideband EoR experiments and present here the arguments that led to the adoption of the SARAS 2 antenna configuration. 

To avoid coupling of sky spatial structure to spectral domain, we consider the class of frequency independent antennas. These are generally based on self-scaling behavior. If the physical dimensions of the antenna are scaled, then its properties do not change if the operating frequency also scales by the same factor. It has been shown that if the shape of the antenna could be specified entirely by angles, its performance would be frequency independent \cite{1150565}.  

Wideband spiral antennas are an example of this class.  However, even if the structural bandwidth well exceeds the operating band, the inevitable truncation of structure at both top and bottom causes reflection of currents, leading to frequency dependence in the beam pattern. Further, if the arms of the spiral are not electrically balanced, the beam would have a squint that rotates with frequency, which would introduce spectral ripples in response to sky structure. Spiral antennas are sensitive to circular polarization.

Linear log-periodic dipoles are the corresponding frequency-independent antennas for linear polarization. They are not strictly frequency independent since their properties in terms of beam pattern and impedance have a periodicity that depends on logarithm of frequency. This periodicity across the band can result in additional frequency dependent structures in the spectrum, particularly for wide bandwidths that are critical for CD/EoR global signal detection.

Another argument against the above categories is that wideband spirals and log-periodic dipoles are electrically large and hence their reflection efficiency will be complex function of frequency.  

Sidelobes, chromatic beams as well as complex reflection efficiencies may be avoided by using electrically small antennas whose physical dimensions are much smaller than the minimum wavelength under consideration. However electrically small antennas are difficult to match to a load due to its low input resistance and high reactance. This results in a low efficiency for short antennas. There is thus a compromise between efficiency and frequency independent performance. An approach is to accept a lower efficiency at long wavelengths since the sky is very bright at long wavelengths and it is adequate to have an efficiency which ensures that sky signal dominates the system temperature.

The short dipole antennas seem to be an attractive choice for CD/EoR detection. However, one of the major concerns of employing dipoles is the use of baluns as discussed in Sec.~\ref{sec:resis}, which leads to a frequency dependent resistive loss that is difficult to characterize. Further, the configuration in which the antenna is used can affect its achromaticity. For example, if the dipole is mounted a certain distance above a conducting plane, there would be multipath propagation of radiation from any sky direction to the dipole - one direct path and a second reflected off the plane.  The relative phase would be frequency dependent and hence the beam pattern of the dipole above reflecting plane would be frequency dependent.   A way to avoid this may be to use absorbers below the antenna to suppress the reflected component. Assuming a sky brightness of a few hundred Kelvin, the absorbers would require to have a power reflection coefficient less than $-100$~dB over the whole band in order to keep any frequency structure in the spectral response below about 1~mK. Absorbers with such specification over 40-200 MHz, implying a bandwidth of 5:1, are impossible with present technology as far as we know.  Additionally, any non-smooth frequency characteristics of the absorption would lead to a non-trivial frequency dependent bandshape for the antenna transfer function.

Short monopole antennas are suitable candidate antennas for CD and EoR detection since they do not require baluns, and beam chromaticity due to multipath propagation can be avoided since there is no physical distance between antenna and ground, latter being part of the antenna. With the absence of a balun or impedance transformer, we do compromise on the antenna efficiency; however, this is the trade off that may be accepted in order to gain a maximally smooth antenna reflection efficiency and reflection coefficient $\Gamma_c$. 

The shape of the monopole radiating element also plays a crucial role in determining the spectral shape of reflection efficiency. Any sharp edges or truncation of the structures, like in a discone antenna \cite[Chapter 7]{StutzmanThiele201205}, will lead to reflection of currents that may interfere to produce complex frequency structure in the impedance characteristics. 

\begin{figure}[htbp]
\begin{center}
\includegraphics[scale=0.5]{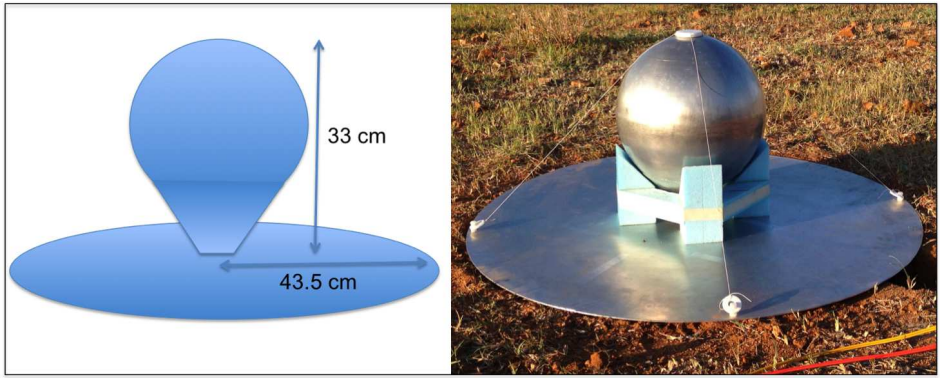}
\caption{SARAS 2 antenna}
\label{fig:1}    
\end{center}   
\end{figure}

We thus choose a sphere-disc type of monopole antenna as the base for the design for SARAS 2, since such an antenna type may be described by a minimum number of parameters. SARAS 2 antenna consists of two elements: a circular aluminium conducting disc on the ground and above that is a sphere that smoothly transforms into a truncated inverted cone as shown in Fig.~\ref{fig:1}. The receiver electronics is mounted beneath the metallic disc, and the vertical coaxial cable connected to the receiver has a central conductor that connects to the vertex of the inverted cone and an outer conductor that connects to the disc. 
The height of the spherical radiating element of the antenna was optimized to be $33~\rm cm$, and the radius of the disc to be 43.5~cm, using WIPL-D electromagnetic simulations.  The optimization aimed at keeping the resonant frequency outside the band, the reflection efficiency maximally smooth and the beam patterns frequency independent, while also striving to maximize the reflection efficiency at low frequencies.  The optimization makes the height of the monopole element less than $\lambda/4$ at the highest frequency, making it electrically short at all operating frequencies.  These aspects are discussed further below.  

\subsubsection{Beam Pattern of the SARAS 2 antenna}

Field measurements of the radiation pattern were made across the band, and these were compared to those derived from electromagnetic simulations of the antenna. 

\begin{figure}[htbp]
\begin{center}
\includegraphics[scale=0.5]{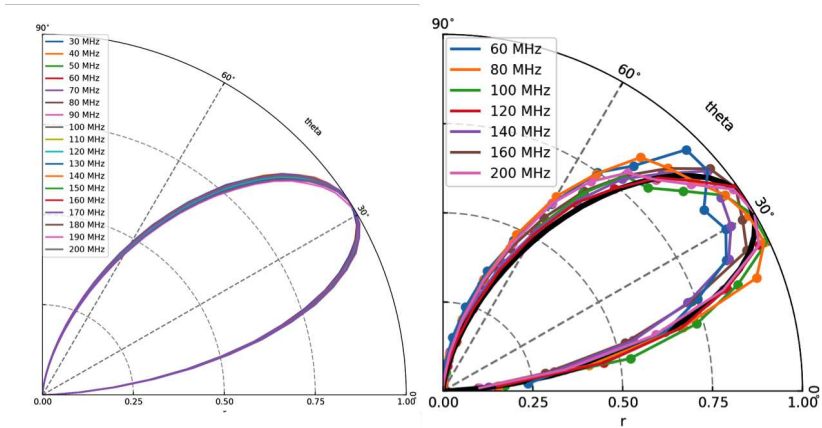}
\caption{The left panel shows the beam pattern at different frequencies as obtained from electromagnetic simulations. The right panel shows the measured beam pattern together with the mean of the simulations shown overlaid as a solid black line.}
\label{fig:2}    
\end{center}   
\end{figure}

A half-wave dipole was used as the transmitter for the measurement; this was separately tuned for measurements at different frequencies. It was clamped $8~\rm m$ above the ground to minimize interactions with the ground, particularly at low frequencies where the wavelength is a few meters. The SARAS 2 antenna was kept stationary and the dipole moved to a set of distances to measure the beam versus elevation angle.  The SARAS 2 antenna was used as the receiving element and the power received, after corrections using the Friis equation \cite[Chapter 2]{Balanis:2005:ATA:1208379}, was used to compute the beam pattern at different frequencies. The simulated and measured beam pattern at different frequencies are shown in Fig.~\ref{fig:2}. 
The measurements, which have measurement accuracy of about $10\%$, agree with the simulations, confirming that the antenna has a frequency independent beam as expected for an electrically short monopole.  It has a maximum response at $30^{\circ}$ elevation from horizon and gradually decreases to zero towards horizon and zenith. The beam is omni-directional implying that it has a non-directional response along azimuth and a directional pattern along elevation with a half-power beam width of $45^{\circ}$.  

\subsubsection{Reflection Efficiency}
\label{label:ref_eff_gamma}

\begin{figure}[htbp]
\begin{center}
\includegraphics[scale=0.6]{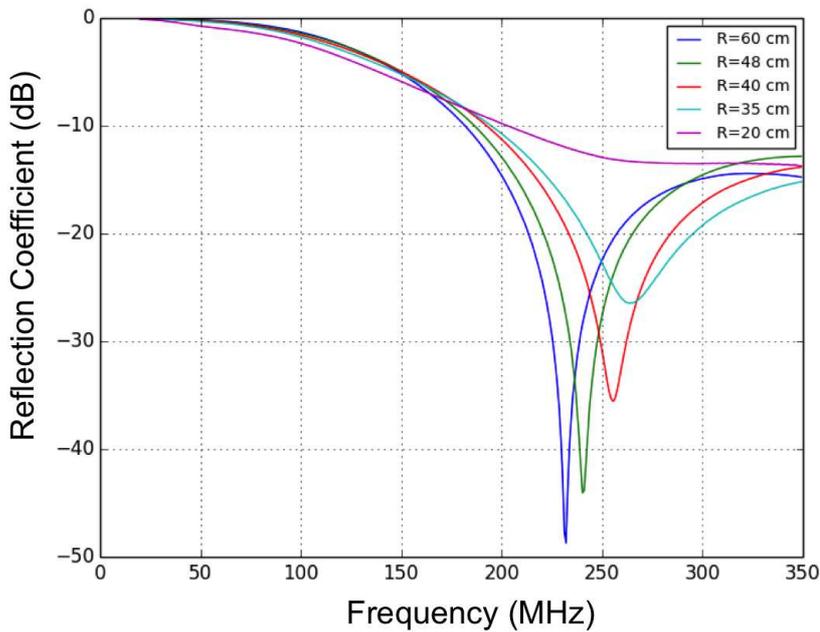}
\caption{Reflection Coefficient versus frequency for different disc radii, for fixed sphere radius.}
\label{fig:resonance}    
\end{center}   
\end{figure}

As a primary consideration, antennas with smoothly varying reflection efficiency, and also maximally smooth $\Gamma_c$, are preferable. For this reason, we avoid any resonance in the band since that would result in a sharp variation of $\Gamma_c$ in the frequency domain at the resonant frequency. The resonant frequency depends critically on the dimensions of the sphere and to a lesser extent on that of the metallic disc below. We have made field measurements of the reflection coefficient with different radii for the metallic disc. As shown in Fig.~\ref{fig:resonance}, shorter the radius of the disc, higher is the resonant frequency, and this is favorable in terms of spectral smoothness since the rate of variation of the reflection coefficient in the band would be slower. However, very small dimensions of the disc would lead to reduction in radiation efficiency.  

We chose the the radius of the disc of the SARAS 2 antenna to be a compromise between sensitivity and spectral smoothness. The disc has radial extent of 0.435~m; thus the antenna operates well within the first resonance, which lies at 260~MHz. Since the reflection of currents from the edge of the disc, for the chosen radial extent, can only result in $\sim 350~\rm MHz$ scale sinusoidal frequency-domain structure in $\Gamma_c$, we do expect $\Gamma_c$, and hence reflection efficiency, to be spectrally smooth since it would contain only about a half of a sinusoidal ripple in the 160~MHz band ($40-200~\rm MHz$).

An alternate approach is to separately measure $\Gamma_c$, perhaps {\it in situ}, and use it to model the data. However, that requires a high precision measurement, close to 1 part in $10^4$, for controlling the systematics to be below a mK (the rationale for this specification has been explained below in Section~\ref{sec:imp_mis}). Measurement at this precision is challenging as the components in the measurement setup itself; {\it e.g.}, any interconnecting cable between the measuring instrument and antenna, may introduce a spurious shape in the reflection coefficient measurement that is not intrinsic to the antenna. Though such cables may be calibrated as part of measurement process, a change in their warp or a small change in impedance due to temperature change or even switching of connectors to make this measurement may render the calibration solution for the measurement inaccurate. 

\begin{figure}[htbp]
\begin{center}
\includegraphics[scale=0.48]{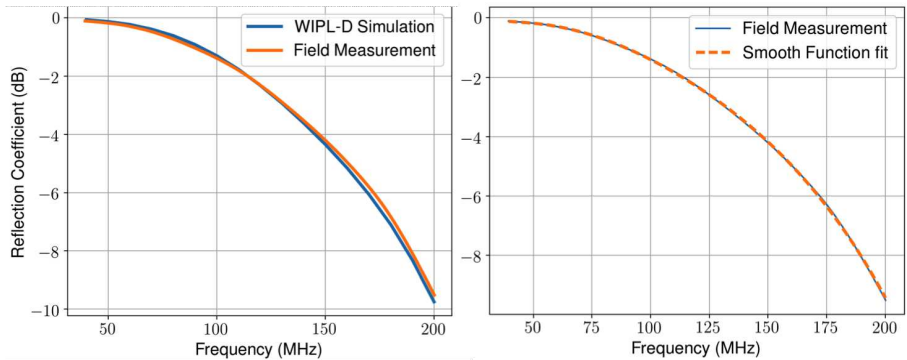}
\caption{The panel on the left shows the measured reflection coefficient for the SARAS 2 antenna overlaid with that from the electromagnetic simulations. The panel on the right shows the maximally smooth function fit to the measured reflection coefficient.}
\label{fig:3}    
\end{center}   
\end{figure}

We made a measurement of the reflection coefficient of the SARAS 2 antenna with extreme care using a rugged field spectrum analyzer, which was placed underground just beneath the antenna and directly connected to the antenna without cables.  The calibration setup and measurement was remotely operated to keep the antenna environment stable during the measurement process. Fig.~\ref{fig:3} shows the measured reflection coefficient, the reflection coefficient expected from electromagnetic simulation, as well as a maximally smooth function fit to the measurement.
The fit residuals show no structure above the measurement noise that is about $10^{-4}$. Thus $\Gamma_c$ has no spectral features to the measured level of accuracy and we discuss below in Section~\ref{sec:imp_mis} the implications for the level of receiver systematics given this upper limit on departures from smoothness in $\Gamma_c$.

Based on the considerations discussed above and also the results of the field measurement, we have adopted an approach in which the measurement data is modeled based on assuming a maximally smooth functional form, with free parameters, for $\Gamma_c$ and hence $\eta_t$.  

\subsubsection{Radiation and Total Efficiency}
\label{sec:tot_eff}

The antenna radiation efficiency can be measured via various methods, e.g. Wheeler Cap, radiometric, directivity/gain method, using waveguides etc. \cite{Huang2014}. Existing methods of efficiency measurement, as described in \cite{1084}, have large errors. Some methods require precisely controlled environmental conditions and anechoic chambers to carry out the measurements and can be time consuming. It is extremely difficult to adopt these methods for the field measurement of $\eta_r(\nu)$ at the accuracy required for the current experiment, which is 1 part in $10^5$. For that reason, we have developed a new method for measuring the total efficiency using the spectral measurements of sky brightness acquired for CD and EoR detection. 

We adopt the GMOSS model \cite{2017AJ....153...26S} as a representation for the sky brightness distribution and compare the absolute sky brightness with measurements made by the spectrometer. The ratio of absolute sky brightness to the foreground estimated from the measurement gives the total efficiency versus frequency. Details of the method are presented in Appendix~\ref{app:1}. Here, in Fig.~\ref{fig:fig_effp}, we present this measured total efficiency. 

\begin{figure}[htbp]
\begin{center}
\includegraphics[scale=0.5]{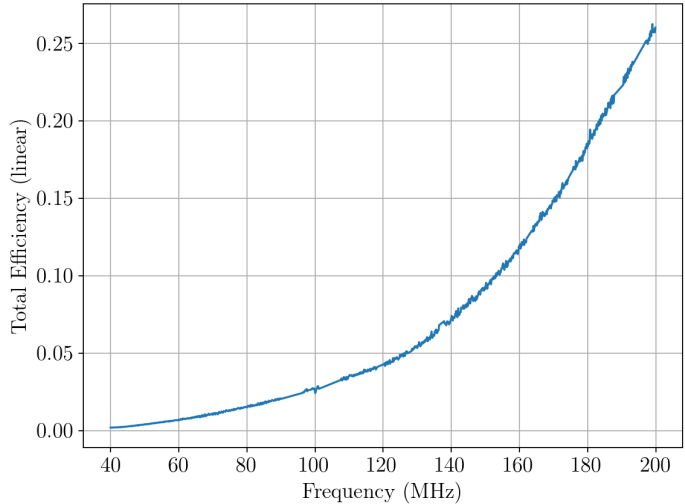}
\caption{Total efficiency versus frequency as derived using the GMOSS model and SARAS 2 measurement data taken during a night.}
\label{fig:fig_effp}    
\end{center}   
\end{figure}

The total efficiency varies monotonically with frequency consistent with a maximally smooth transfer function for the transformation from the sky spectrum to the measurement data. However, the actual magnitude of the efficiency does indeed decrease fairly sharply to a few per cent at low frequencies. Therefore, as mentioned above, we have restricted the analysis of SARAS 2 data to above $110~\rm MHz$, marking the upper end of FM band.

\section{Analog Signal Processing}

The beam-weighted sky signal is coupled into the system with a multiplicative gain, $\eta_t$, that is the total efficiency of the antenna. Additionally, the signal further undergoes a multiplicative gain, which we refer to as bandpass, arising from the gains of the devices in the receiver after antenna. The receiver, which follows the antenna in the signal path, is designed with the following considerations:

\begin{itemize}
\item The receiver requires a calibration scheme by which the bandpass, which is a multiplicative gain factor for the measurement data, may be flattened.
\item Unwanted additive spurious, contributed by the receiver, needs cancellation or a method by which they do not confuse any CD/EoR detection.
\item The receiver chain is designed to distribute the gains and the resulting power levels along the signal path to maintain linearity and low levels of intermodulation products.
\end{itemize}

\subsection{Calibration considerations}
\label{sec:S_recv}

The antenna signal entering the receiver is modified by the receiver gain. The spectral behavior of system gain is the cumulative product of individual gains of all the modules that the signal passes through. We term the process of correcting the measurement data for frequency dependent multiplicative gain and hence flattening the instrument spectral response as bandpass calibration. At the same time, the arbitrary counts in which measurement data is acquired needs to be scaled to be Kelvin units of antenna temperature. This is termed as absolute calibration. It is more critical for a global CD/EoR experiment to attain a high precision for bandpass calibration---so that residual errors are within about a mK---compared to getting the absolute temperature scale right. This is so since the latter is simply a scaling factor to the data while an erroneous bandpass calibration can potentially distort the shape of the spectrum. This is a major consideration for the receiver design.

Bandpass calibration, along with calibration for $\eta_t$, can be achieved via various means. One way is to have a spectrally flat, broadband signal external to the antenna that traverses the same path as that of the sky signal. It would then be able to remove the frequency structure imposed by $\eta_t$ as well as by the system bandpass, given its intrinsic spectral flatness. Such a calibration signal is required to be externally generated by a transmitting system.  The difficulty with such an approach is that the problem of bandpass and $\eta_t$ calibration is not actually solved but simply transferred to the transmitting system!  

An alternate approach to generating an external flat-band signal is to deploy a pulse calibration scheme \cite{2017ExA....43..119P}.  The method involves generating short duration pulses, with time domain width substantially smaller than the bandwidth of the receiver, and using the measurement data to compute and correct the bandpass and $\eta_t$.  The disadvantage of this approach is that in order to attain adequate signal-to-noise for the calibration, the short duration pulses are required to be of high amplitude, which requires a high dynamic range receiver.

The external calibrator source could be an astronomical source like Cas A, the Moon, etc. \cite{1999A&A...345..380S}. Astronomical sources for calibration have been routinely used in interferometer measurements \cite{Perley_Schwab_Bridle_1989}. Although astronomical continuum sources may have spectrally smooth emission over the bands of interest here, the spectrum of the Moon may be corrupted by reflected Earthshine, particularly in the FM band \cite{2015MNRAS.450.2291V}. The primary argument against using astronomical sources for CD/EoR radiometer calibration is that the antennas used for such experiments have a small effective collecting area, and even the brightest of point sources would contribute only a few Kelvin in antenna temperature for this class of antennas. Consider, for example, Cas A, which is one of the brightest point sources in the long wavelength radio sky. Its flux at $150~\rm MHz$ is $\sim 11.7~\rm kJy$ \cite{1977A&A....61...99B}.  The antenna temperature due to a point source in the sky is given by \cite{9780691137797}:
\begin{equation}
T_A = \frac{A_e S}{2k},
\end{equation}
where $A_e$ is the effective collecting area of the antenna, $S$ is the flux of the source and $k$ is the Boltzmann constant. The effective area of a monopole antenna at this frequency would be close to $1~\rm m^2$ \cite{9780824704964}.  We infer that even for an extremely strong celestial source like Cas~A, $T_A$ is $\sim 4.5~\rm K$. Thus, compact strong celestial sources are not suitable candidates for calibration, since the temperature increment when the source comes into the beam would be significantly smaller than the system temperature, which is usually at least a few hundreds of Kelvin. 

A more attractive solution to the calibration problem for CD/EoR radiometers is the use of a broadband noise source internal to the system, where there is a better control over the spectral flatness of the signal injected into the signal path.  Internal calibration sources may also be switched with small duty cycles and so the calibration may be performed in shorter time intervals thus accounting for shorter period temporal variations in the bandpass. However, since the calibration signal is injected into the signal path after the sky signal has been coupled into the system through the antenna, the characteristics introduced by the antenna, $\eta_t$, cannot be removed by such calibration and hence needs to be modeled separately. 

Further, there are choices the way the calibration signal is coupled into the signal path. A widely used scheme is Dicke switching \cite{Dicke1982} where a switch is used to swap the receiver input between the antenna and the noise source. The spectra obtained in the two switch positions are subtracted to derive a gain solution which is applied to the data \cite{2012RaSc...47.0K06R}. However, the receiver noise related additives appearing in the measurement data in the two switch positions may differ due to different impedance characteristics of the antenna and the noise source. Thus the subtraction of the two spectra would create another frequency structure in the calibration solution that can be difficult to model. An alternate strategy is to have a method in which the antenna and noise source are both always connected to the system, and the noise power level of the calibration source is switched between high and low states, so that the nature of internal systematics does not alter in the process of calibration. We explore this approach further in Sec.~\ref{sec:recv_arch} below, where we describe the SARAS 2 receiver architecture.

Maintaining linearity in the signal path is important for any of the above calibration schemes to work.  As mentioned above, this requires that while power levels are maintained to be considerably above the noise floor of the system so that there is no degradation of signal-to-noise ratio along the signal path, at the same time sufficient headroom is maintained between the operating power and saturation limits.  

\subsection{Considerations related to additives from receiver noise}
\label{sec:S_recv}

Another parameter which plays an important role in deciding the architecture of the analog receiver is the spectral behavior of the additive arising as a result of multi-path propagation of noise from the Low-Noise Amplifiers (LNAs), which propagates in forward and reverse directions. Since the antenna and LNA impedances are not perfectly matched along with their interconnect, a part of the noise from the LNA that travels towards the antenna is reflected back.  Interference between this reflected component and the forward propagating receiver noise results in a systematic additive in the measurement data \cite{meys1978wave}. This multi-path propagation of receiver noise voltages and their addition results in a sinusoidal variation for the systematic additive versus frequency, which is also modulated by the spectral shape in $\Gamma_c$. The period of the sinusoid is governed by the phase difference between the interfering components and hence on the length of the system between the impedance mismatches on the two sides of the LNA. The amplitude of the response depends on absolute values of $\Gamma_c$, noise figure of LNAs and the magnitude of correlation between the forward and back-propagating components of LNA noise. 

Thus, an important criterion in receiver design is to minimize the amplitude and shape the spectral behavior of this additive receiver response to be maximally smooth. It gains importance due to the fact that this internal receiver related component in the measurement data is an additive and is often not calibratable. Various system design considerations can make this receiver additive spectrally smooth. First, the electrical length of the analog receiver chain can be made so short that the period of the sinusoid is significantly larger than the band of operation. This would ensure that only part of a cycle of the sinusoid appears in the full band and hence appears smooth. Second, as discussed below in Sec.~\ref{sec:imp_mis}, maintaining $\Gamma_c$ to be spectrally maximally smooth is an additional way to keep this component devoid of complex spectral features.  The amplitude of this additive can be further reduced by using LNAs with low noise figure, by making the antenna impedance better matched to that of receiver and thereby lowering the value of $\Gamma_c$, and by selecting LNA designs that reduce the correlation between the forward and reverse traveling LNA noise components. 

\subsection{The SARAS 2 receiver}
\label{sec:recv_arch}

\begin{figure}[htbp]
\begin{center}
\includegraphics[scale=0.5]{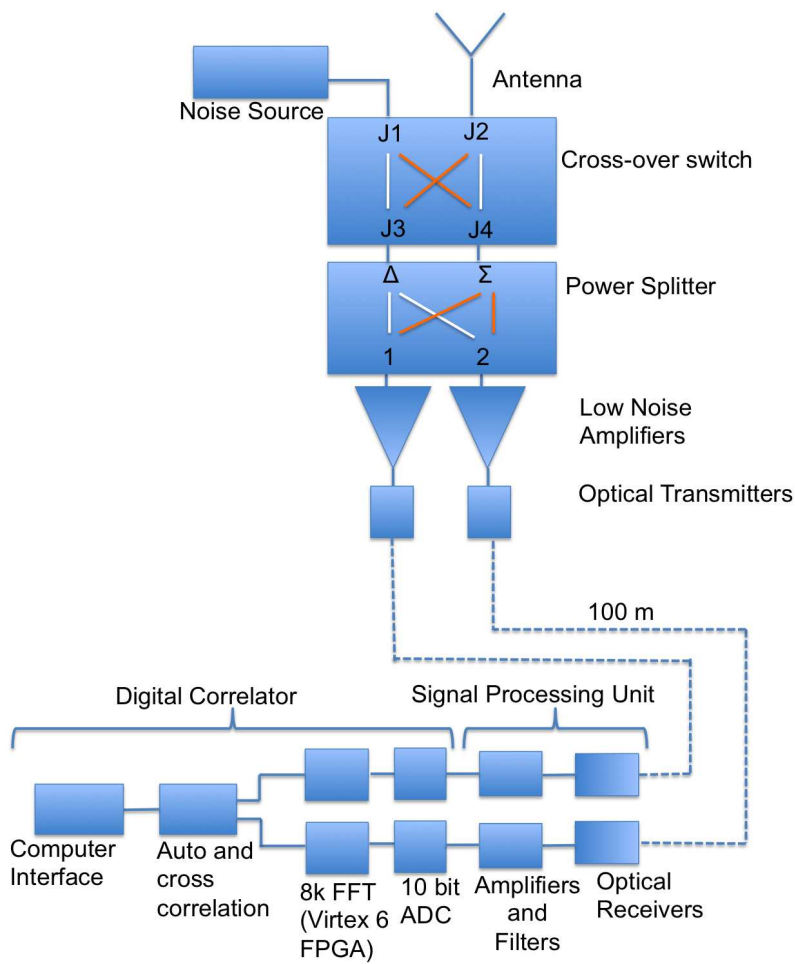}
\caption{Schematic of the SARAS 2 receiver.}
\label{fig:recv}    
\end{center}   
\end{figure}

The SARAS 2 receiver uses an internal noise source for generating the calibration signal which is connected to a four port cross-over switch as shown in Fig.~\ref{fig:recv}. When the noise source is in OFF state, this device serves as a reference for the measurement of the sky signal.  The antenna is connected to the other input of the switch.  The outputs of the switch go to a power splitter module that provide the sum and difference of the two inputs to two paths of the receiver chain. Hereinafter we refer to the two analog signal paths beyond the power splitter as the two arms of the analog receiver. The switch swaps the antenna and reference/calibration signals between the two ports of the power splitter, so that the receiver arm picking up the difference signal alternately gets sky minus reference and reference minus sky.  For each position of the switch, the noise source is switched on and off with a cadence of about one second.  SARAS 2 thus cycles over four system \textit{states} as shown in Table~\ref{tab:1}.  

\begin{table}
\caption{States of the system}
\label{tab:1}       
\begin{tabular}{lll}
\hline\noalign{\smallskip}
State & Noise Source & Switch Position \\
\noalign{\smallskip}\hline\noalign{\smallskip}
OBS0 & OFF & 0 \\ 
CAL0 & ON & 0 \\
OBS1 & OFF & 1 \\
CAL1 & ON & 1 \\ 
\noalign{\smallskip}\hline
\end{tabular}
\end{table}

The analog receiver is powered by a Li-Ion battery pack that is mounted with the receiver, in a metallic enclosure, beneath the metallic disc of the antenna. By this we avoid any conductive power lines running external to the antenna, which may result in unwanted coupling between the electromagnetic field in the neighbourhood of the antenna and the signal path within the receiver.  Further, such an arrangement is essential to be able to deploy the antenna in remote locations where external power is not available. 

\subsubsection{Signal Flow}

We refer to the signal that is coupled to the receiver via the antenna as $T_A$. We denote the components of the measurement data arising from the noise source in ON and OFF states as $T_{\rm CAL}$ and $T_{\rm REF}$ respectively. $T_{\rm REF}$ is the power from a reference termination---a well matched precision $50~\mathrm{\Omega}$ termination---that will correspond to a noise temperature of value of the ambient temperature, which is approximately $\sim 300~\rm K$. The two signals, from the antenna and from the reference/calibration, are inputs to the cross-over switch, as shown in Fig.~\ref{fig:recv}.  When the switch is in position $``0"$, the signal at the input port J1 of the cross-over switch is channeled to output J3. Similarly the signal at input J2 appears at output J4. In position $``1"$ of the switch, the paths are crossed implying that the signal at J1 goes to J4 while that at J2 appears at J3. The two switch positions are denoted in two colors in Fig.~\ref{fig:recv}.

The pair of signals are then fed to pair of input ports of a power splitter---the sum port $\Sigma$ and difference port $\Delta$---depending on the switch state. The signals undergo a voltage attenuation ($g$) while passing through the splitter. The signal at the $\Sigma$ port appears at both outputs of the splitter in phase while the signal at the $\Delta$ port appears at the two outputs with a phase difference of $180^{\circ}$. 

The signals are transmitted from the receiver to the signal processing unit as RF over fiber: as analog signals modulating the intensity of laser light.  Demodulation of the optical signal at the signal processing unit gives back the RF signal. This unit, along with the following digital correlator, are placed about $100~\rm m$ away from the antenna in an electromagnetically shielded environment to avoid any electromagnetic interference being picked up by the antenna. 

The signals in the entire band are lowpass filtered at the signal processing unit such that frequencies above $230~\rm MHz$ are filtered out. It may be noted here that the amplifiers, attenuators and indeed all components used in the two arms within the signal processing unit do not contribute to any additive in the final measurement data, because the arms are optically isolated from each other and hence signals from one arm do not couple to the other. The signals in the pair of receiver arms finally enter the digital spectrometer where they are digitized, Fourier transformed and cross-correlated to produce the measurement data. 

In the SARAS 2 receiver, the sky power, after being modified by the total efficiency, is $\sim 300~\rm K$ which corresponds to a power of $~\sim-90~\rm dBm$ in the band of $40-200~\rm MHz$. The gain at each stage within the receiver arm is chosen such that all devices operate well below saturation and continue to be in linear regime of operation.  This criterion becomes more stringent farther in the signal path where power levels progressively increase with each amplifier stage. The gains in the system have been adjusted such that the input power at the last amplifier of the signal processing unit is $-52~\rm dBm$.  This amplifier is chosen to have a 1~dB compression point at +22~dBm; therefore the operating level even at this most critical stage is about 74~dB below saturation. 

\subsubsection{Bandpass Calibration} 
\label{sec:BP_cal}

We now write expressions for the power measured, in temperature units, in different states of the system. Since SARAS 2 is a correlation spectrometer,  the mathematical operations performed here are complex operations. The subscripts OBS0, CAL0, OBS1 and CAL1 represent the system states as listed in Table~\ref{tab:1}.
\begin{align}
\label{eq:diff1}
T_{\rm OBS0} & =  G_1G_2^* g^2(T_A - T_ {\rm REF} ) + P_{\rm cor}, \\
T_{\rm CAL0} & = G_1G_2^* g^2(T_A - T_ {\rm CAL}) + P_{\rm cor}, \\
T_{\rm OBS1} & = -G_1G_2^* g^2(T_A - T_ {\rm REF}) + P_{\rm cor}, \ {\rm  and} \\
T_{\rm CAL1} & = -G_1G_2^* g^2(T_A - T_ {\rm CAL}) + P_{\rm cor},
\end{align}
where $G_1$ and $G_2$ are the gains in the two receiver arms and $P_{cor}$ is the unwanted power appearing in the measurement data due to any spurious coupling of signals between the two arms either within the signal processing unit or at the samplers. $P_{cor}$ would not be expected to change in magnitude or phase in different states and hence subtracting any pair of measurement data would cancel this additive.  With this aim, we difference the measurements in the two switch states that have the same state of the noise source. We thus get two differential spectra:
\begin{align}
T_{\rm OFF} &  =  T_{\rm OBS0} - T_{\rm OBS1} \nonumber \\
& = 2G_1G_2^* g^2(T_A - T_ {\rm REF} ), \  {\rm and}
\label{eq:BP1} \\
T_{\rm ON} & = T_{\rm CAL0} - T_{\rm CAL1} \nonumber  \\
& = 2G_1G_2^* g^2(T_A - T_ {\rm CAL} ).
\label{eq:BP2} 
\end{align}
We next derive a measure of the system bandpass by differencing the two spectra computed above in Eq.~\ref{eq:BP1} and \ref{eq:BP2}:
\begin{align}
T_{\rm TEMP} & = T_{\rm ON} - T_{\rm OFF} \nonumber \\
& = -2G_1G_2^* g^2(T_ {\rm CAL} - T_ {\rm REF}).
\label{eq:TEMP}
\end{align}

This complex spectrum represents the system bandpass, which we we use to calibrate the measurement data for the bandpass.  The term $(T_ {\rm CAL} - T_ {\rm REF})$ represents the \textit{step} change in the noise temperature from the reference port when the noise source is switched on, and is the excess power above the OFF state.  This step in power is also referred to as the Excess Noise Ratio (ENR) of the noise source; we call this $T_{\rm STEP}$.

We divide Eq.~\ref{eq:BP1}, which represents the measurement data with calibration source off, by Eq.~\ref{eq:TEMP}, which represents the bandpass calibration, to flatten the system bandpass:
\begin{align}
\label{eq:abs_cal}
\frac{T_{\rm OFF}}{T_{\rm TEMP}} = -\frac{(T_A - T_ {\rm REF} )}{T_{\rm STEP}}.
\end{align}
This calibration, being a complex division, also results in the sky data being in the real component of the complex calibrated spectrum and yields the differential measurement:
\begin{align}
\label{eq:abs_cal_2}
T_A - T_ {\rm REF} = -\frac{T_{\rm OFF}}{T_{\rm TEMP}}{T_{\rm STEP}}.
\end{align}

This signal processing cancels any internal additive systematics originating in the signal processing unit and digital signal processor, as shown in the process of differencing spectra through Eq.~\ref{eq:diff1} - ~\ref{eq:BP2}, and also performs a complex bandpass calibration of the measurement data, as shown in Eq.~\ref{eq:abs_cal_2}. We finally get a differential measurement of the antenna temperature $T_A$ with reference to the termination $T_{\rm REF}$. The only unknown is the power step corresponding to the difference in the noise source in ON state compared to OFF state, $T_{\rm STEP}$. We discuss the method adopted to derive the value of $T_{\rm STEP}$ next.

\subsubsection{Absolute Calibration}
\label{sec:abs_cal}

Absolute calibration for the measurement data is provided by determining the scaling factor for the data from the arbitrary counts in which data is acquired to Kelvin units.  $T_{\rm STEP}$ is used in Eq.~\ref{eq:abs_cal_2} above to convert the calibrated spectra from arbitrary units to Kelvin units. 

In order to measure this temperature step for the calibration, we make a laboratory measurement using the receiver.  The antenna is replaced with a precision $50~\mathrm{\Omega}$ termination. Temperature probes are firmly fixed on the outer conductor of this termination and another on the reference termination. We now immerse the termination that is in place of the antenna, along with its temperature probe, into hot water in a thermally insulated dewar and let it cool slowly over time. The temperatures of the terminations are logged by the probes. At the same time, the bandpass calibrated power is recorded by the receiver system. We repeat this exercise by immersing the termination that is in place of the antenna in ice water and let this bath heat slowly over time to ambient temperature.

We denote the true physical temperatures of the termination and reference loads by $T_{A}$  and $T_{\rm REF}$, while their respective temperatures as measured by the probes are denoted by $T_{a_m}$ and ${T_{r_m}}$.  Considering the reference load, its true temperature $T_{\rm REF}$ would always be somewhat higher than the measured $T_{r_m}$ since the measurement from the probe is on the outer conductor of the probe which would be cooler than the actual temperature. Thus, we may write that
\begin{equation}
T_{\rm REF} = T_{r_m} + k_1,
\end{equation}
where $k_1$ is always positive. Similarly, for the termination that replaces the antenna, we have
\begin{equation}
T_{A} = T_{a_m} + k_2,
\end{equation}
where $k_2$ can both be positive or negative. When the termination is immersed in hot water,  $T_{a_m}$ would overestimate the true temperature of the termination where as when immersed in ice water bath, it would be lower than $T_{A}$. Both these effects are due to the fact that there is thermal resistance between the outer metallic bodies of the terminations, where the temperature probes are fastened, and the source of electrical noise is at the core of the electrical resistance within the terminations. Hence the probe measurement either leads or lags depending on whether the termination is placed in an environment that is above or below the ambient temperature respectively.

Thus, from the experiments with hot and cold water baths we have two sets of physical temperature measurements for the LHS of Eq.~\ref{eq:abs_cal_2} and corresponding  ratios ${T_{\rm OFF}}/{T_{\rm TEMP}}$ from corresponding measurement data.  A  plot of the difference of the two temperature probes versus the corresponding ratios from the measurement data is expected to result in a straight line, with the slope of the line yielding $T_{\rm STEP}$ in accordance with Eq.~\ref{eq:abs_cal_2}.  We also solve this data for offsets $k_1$ and $k_2$ in the straight line model to account for the difference between the measured and true temperatures in each of the temperature probes. 

\begin{figure}[htbp]
\begin{center}
\includegraphics[scale=0.3]{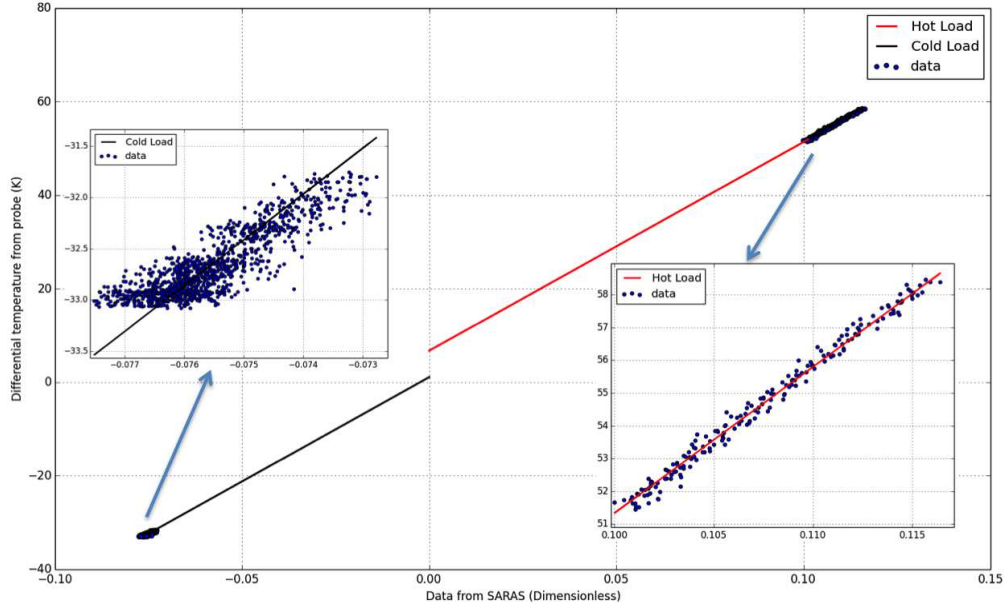}
\caption{Fit that yields the Absolute Calibration Scale factor $T_{\rm STEP}$.}
\label{fig:abs_cal}    
\end{center}   
\end{figure}

We thus plot the probe measurements versus the system measurements and model each of the hot and cold bath experimental data as straight lines, constraining the slopes to be same, but allowing for different intercepts. The common slope gives an estimate of $T_{\rm STEP}$ of value $ 446~\rm K$ with $1\%$ accuracy. We show the data and their model fits in Fig.~\ref{fig:abs_cal}. 

\subsubsection{The measurement equation}
\label{sec:imp_mis}

There are three sources of signals within the system:
\begin{itemize}
\item sky and ground radiation entering through the antenna, resulting in an antenna temperature $T_A$,
\item signal from the reference termination $T_{\rm REF}$, which becomes the calibration signal $T_{\rm CAL}$ when the calibration source is on, and
\item signals corresponding to receiver noise from the LNAs, corresponding to the receiver noise temperatures $T_{N_1}$ and $T_{N_2}$ that are the noise figures of the LNAs.
\end{itemize}

Since the antenna and the LNA's have impedances at their ports that are not perfectly matched to the interconnects, all of the above signals propagating along interconnects get partially reflected at their terminals.  All these signals thus suffer multipath propagation with differential delays from their respective sources to the digital signal processor and interfere to produce frequency dependent shapes. However, owing to the correlation spectrometer scheme adopted, a significant part of receiver signal arriving at the correlator from the two arms is uncorrelated and does not result in any response.  It is when receiver noise signal from an LNA in one receiver arm propagates to the antenna and reflects back into the other receiver arm that we have a correlated receiver response in the measurement data. Thus the amplitude of receiver related component in the measurement is reduced relative to that in autocorrelation spectrometers.  The formalism and derivation of additives arising due to impedance mismatch in correlation spectrometers, for the SARAS 1 system, is in \cite{2013ExA....36..319P}.   The configuration in SARAS~2 is somewhat different from that in SARAS~1 and we provide below generalized expressions for the calibrated measurement data, with multi-order reflections, without pedagogical derivation.
\begin{align}
T_{\rm meas} & = \left[ \left(\frac{C_1}{C_2}\right) T_A -  T_{\rm REF} + \left(\frac{C_{n1}}{C_2}\right) T_{\rm N_1} + \left(\frac{C_{n2}}{C_2} \right) T_{\rm N_2} \right]T_{\rm STEP},\  \textrm{ where}
\label{eq:meas} \\
C_1 & = \left[ \sum\limits_{l=0}^{\infty} |\gamma^{2l}| \sum\limits_{m=0}^{\infty}\Re(\gamma^m\mathrm{e}^{i m\phi}) \right] ,
\label{eq:c1} \\
C_2 & = \left[ 1 - |\psi|^2 \left( \sum\limits_{l=0}^{\infty} \gamma^l \mathrm{e}^{i (l+1)\phi} \right) \left( \sum\limits_{m=0}^{\infty}\gamma^m \mathrm{e}^{i (m+1) \phi}\right)^{*} + 2 i \Im \left\{ \psi \left( \sum\limits_{n=0}^{\infty} \gamma^n \mathrm{e}^{i (n+1)\phi} \right) \right\} \right],
\label{eq:c2} \\
C_{n1} & = f_1 \chi^* + f_1^2 |\chi|^2, \textrm{ and} \\
C_{n2} & = f_2 \chi    + f_2^2 |\chi|^2.
\label{eq:cl} \\
\end{align}
The expansions for $\gamma$, $\psi$ and $\chi$ are:
\begin{align}
\gamma & = (\Gamma_1 + \Gamma_2)\Gamma_a g^2, \\
\psi & = (\Gamma_1 - \Gamma_2)\Gamma_a g^2, \textrm { and} \\
\chi & = g^2 \Gamma_a \mathrm{e}^{i\phi}\sum\limits_{l=0}^{\infty}\gamma^l \mathrm{e}^{i l \phi}.
\end{align}

The term $C_1$ in the RHS of Eq.~\ref{eq:meas} represents the antenna signal and its associated reflections at the LNA, and $C_2$ refers to the signal from the reference termination and its reflections at the LNA and the antenna terminal. For each of these components, the terms in Eq.~\ref{eq:c1} and \ref{eq:c2} correspond to the response due to direct propagation of the signals along the two receiver arms, propagation along the two arms with multiple internal reflections but with equal delays in both arms, and lastly propagation along the two arms and arriving at the digital signal processor with unequal delays. 

The last two terms in Eq.~\ref{eq:meas} represent the response to receiver noise signals that arise from the interference of forward and reverse propagating noise from the individual LNAs that arrive at the digital signal processor along the two arms. $\Gamma_1$ and $\Gamma_2$ are the reflection coefficients at the inputs of the LNAs, and $f_1$ and $f_2$ are the respective correlation coefficients between the forward and reverse traveling components of receiver noise voltages of the two LNAs. $\Phi$ is the phase difference between the forward and reflected signals, which depends on the phase difference due to the length of the system as well as the additional phase shift introduced by reflection $\Gamma_c$.

In ideal conditions, where the antenna and LNAs are perfectly matched with the rest of the system, only the direct path would exist resulting in $C_1$ and $C_2$ to be unity and $C_{n1}$ and $C_{n2}$ to be zero. In such a case, $T_{meas}$ would simply be $T_{A} - T_{\rm REF}$ as given in Eq.~\ref{eq:abs_cal_2}.

In order to minimize spectral variations in these terms, we have miniaturized the overall physical length of the system to reduce the impact of the phase terms that result in sinusoidal responses in frequency.  The total path length was reduced so that the period of the ripple increased and hence the observing band has only a fraction of a sinusoid, thereby maintaining smoothness in responses to the above sources of signals in the system. Through the choice of broadband LNAs, we expect a minimal variation of $\Gamma_1$ and $\Gamma_2$ across the band of operation. The correlation coefficients $f_1$ and $f_2$ were measured separately using the method described in \cite{2013ExA....36..319P}. They are found to be $\sim 10\%$ for the LNAs in SARAS 2. 

Using these values, the amplitude of receiver response is estimated to be $10~\rm K$, which is multiplied by $\Gamma_c$. Since $\Gamma_c$ is shown to be maximally smooth to at least 1 part in $10^4$ (Sec.~\ref{label:ref_eff_gamma}), any deviation of receiver response from smoothness would at most be at the sub-mK level.

We further remark that our estimates for $\Gamma_1$, $\Gamma_2$, $\Gamma_c$ etc. provide the mechanism to decide on the number of higher orders in Eq.~\ref{eq:c1}~--~\ref{eq:cl} that require to be included in the modeling so that the contribution from unaccounted reflections drops below a $\rm mK$. We expand on this while analyzing test data acquired using precision terminations in Sec~\ref{sec:per_meas}.

\section{Digital Signal Processing}

The digital correlator is the last signal processing section of SARAS~2. This computes the auto-correlation spectra of the signals in the two arms of the receiver and the cross-correlation spectrum between the two arms. The auto-correlation spectrum is a real-valued function of frequency where as the cross-correlation spectrum is a complex-valued function.

The first module in the correlator is an Analog-to-Digital Converter (ADC) that digitizes the two analog signals into 10-bit digital levels with a sampling frequency of $500~\rm MHz$. The signals are then windowed using a four-term Blackman-Nuttall window \cite{1981ITASS..29...84N} and channelized using an 8K FFT algorithm implemented on a Virtex-6 FPGA. The sampling, windowing and Fourier transformation of the time-domain voltage waveforms results in 2048 independent complex numbers, corresponding to complex-valued samples of voltages in a 2048-point filter bank spanning the 0--250~MHz band, in each of the two signal paths. This provides an effective frequency resolution of $122 ~\rm kHz$. These complex outputs of the Fourier transforms from the two arms are used to generate the cross-correlation spectrum as well as auto-correlation spectra for each of the two receiver arms separately \cite{6930031}. These spectra are streamed by the FPGA in the form of data packets to a computer. The data, acquired through User Datagram Protocol (UDP), is then processed to construct the spectra with high fidelity. The spectra are written and stored in hard disk of the acquisition PC in MIRIAD file format \cite{1995ASPC...77..433S}.  While the cross-correlation spectrum is used in the data analysis, the auto-correlation spectra are useful for estimating the spectral power in each analog arm and also serve as a good system diagnostic tool.

For the sensitivity requirements of the present experiment, we now derive tolerances on various aspects of the design and performance of the digital system.

\subsection{Tolerance on the clock jitter}

Jitter in the sampling clock leads to uncertainty in the sampled amplitude of the input signal \cite{5995856}. The uncertainty increases with increase in the frequency of the input signal. This results in a deterioration of the Signal-to-Noise ratio (SNR) in the ADC, which is given by \cite{neu2010clock}:
\begin{equation}
\textrm{SNR}_{\rm jitter} ({\rm in~dBc~units}) = -20 \textrm{log}_{10} ( 2 \pi f_{\rm in} t_{\rm jitter} ),
\label{eq:ADC}
\end{equation}
where $f_{\rm in}$ is the input frequency of the signal and $t_{\rm jitter}$ is the clock jitter. 

The SNR of the ADC is also limited by thermal noise and other spectral components, including harmonics of the input signal \cite{kester2009understand}.  This is quantified as the Signal-to-Noise-and-Distortion (SINAD), which is the ratio of the root-mean-square (RMS) signal amplitude to the mean value of the root-sum-square of noise and all other spectral components.  Thus $\rm SNR_{\rm jitter}$ should be below the SINAD of the ADC so as to avoid any deterioration in total SNR. For the ADC selected for SARAS 2, which is a 10-bit sampler, SINAD is 48.7 dB.  From Eq.~\ref{eq:ADC}, we infer that $t_{\rm jitter}$ should be less than $2.9~\rm ps$ considering operation at the highest frequency of $250~\rm MHz$. The actual jitter in the sampling clock, derived from the SARAS 2 synthesizer, is $1.8~\rm fs$, which is well within the tolerance derived above. 

\subsection{Tolerance on the clock drift}

The sampling frequency of the clock might drift over time and this can lead to inaccuracy in the bandpass calibration. To estimate the tolerance on clock stability, we examine the maximum slope in the total system bandpass.  

The bandshape is found to have a variation of $0.8~\rm dB$ with two cycles of ripples over the band of $40-200~\rm MHz$. Assuming a maximum correlated response of $300~\rm K$, including RFI, foregrounds and system contribution, it would result in an overall ripple of peak-to-peak amplitude $60~\rm K$.   We may model this variation as a sinusoid in frequency domain, given by $T = 30 \textrm{sin}(2\pi \tau \nu)$, where $\tau=1/80~\rm MHz^{-1}$. For a frequency shift of $d \nu$, we estimate the change in the measured temperature to be $\frac{dT}{d\nu} = 2\pi \times 30 \tau \textrm{cos}(2 \pi \tau \nu)$. For the experiment, it is desirable to have ${dT}\leq 1~\rm mK$. This would result in a maximum allowed frequency shift to be less than $424 ~\rm Hz$. Given that the sampling frequency is $500~\rm MHz$, we infer that the tolerance on the fractional frequency stability is $8.5\times10^{-7}$. 

SARAS~2 uses a rubidium oscillator as the primary frequency standard for deriving the sampling clock. There is also an option for GPS disciplining built in for long term stability.  SARAS~2 sampling clock, disciplined by a rubidium oscillator, has a fractional frequency stability of $10^{-10}$; therefore the design fulfills the required tolerance on the clock stability.

\subsection{RFI leakage}

A fraction of the power in any frequency channel leaks into neighboring channels in any filter-bank spectrometer. This is of particular concern when there is RFI and its leakage into neighboring channels results in corruption and hence loss of a large number of channels on either side of the frequency of interference. Although the RFI in the central channel might be detected using algorithms discussed below in Sec.~\ref{sec:software}, their contamination over the spectrum is difficult to estimate at the levels necessary for this experiment.  

SARAS 2 uses a Blackman-Nuttall windowing of the time sequences to suppress the spillover of signals in any frequency channel into adjacent channels. This leads to loss in spectral resolution and also sensitivity by a factor of two; however, the windowing results in modifying the point spread function defining the spectral channels so that sidelobes in the spectral domain are substantially suppressed.

We have measured the suppression factor to be better than $10^8$ in power. Thus even if an RFI in a channel is as strong as $10^5~\rm K$, its contribution in the rest of the independent channels would still be at a mK level. This threshold on tolerable interference sets thresholds for the RFI rejection algorithm in that spectra with RFI exceeding this threshold are completely rejected.  Second, the threshold suggests that the observing site needs to be one in which there is no continuous RFI exceeding $10^5~\rm K$. 

\subsection{RFI headroom}

The gains in the amplifiers of the receiver arms are set so that there is sufficient headroom for RFI and the system continues to operate in the linear regime while experiencing tolerable RFI. At the end of the receiver arm, the input power at the ADC is such that it does not exceed its full scale. This ensures that the signal is not clipped in digital domain even if the total power increases appreciably due to presence of a strong RFI. The SARAS~2 ADC clips if a sinusoid signal input to the device has a power exceeding $-2$~dBm.  The SARAS 2 system presents what is almost always a Gaussian random noise voltage to the ADC, whose power is set be be nominally at a much lower total power of $-28~\rm dBm$, which is $30~\rm dB$ higher than the noise floor of the ADC but is also sufficiently below the clipping level. At this level, the probability of any random sample to be close to the clip level is vanishingly small. This reduces the effective number of bits available for the digitization of the signal; however provides enough headroom for strong RFI. Typically during observing at radio quiet sites, it is very unlikely that RFI increases the total power by even a few dB and, therefore, SARAS 2 is guaranteed to operate without non-linear effects of saturation due to the spectrometer and yield useful data during most of the observing duration.

\section{Algorithms: Calibration and RFI rejection}
\label{sec:software}

In this section we describe the data processing steps that are used off-line on the measurement data acquired. These processing steps primarily cater to the calibration of the data, rejection of RFI, and  computing noise estimates for each frequency channel. These noise estimates differ across the spectrum due to differing number of samples rejected due to RFI and their propagation through the different processing steps.

The cadence in each system state (Table~\ref{tab:1}) is 1~s.  In each state a set of 16 spectral records are acquired, each with integration time of 1/16~s. We refer to the set of 16 spectral records as a frame. Each spectral record consists of a complex cross correlation spectrum, representing the cross correlation between the signals in the two arms of the correlation spectrometer, and their auto-correlation spectra, representing the power spectra corresponding to the signals in each arm. 

In the following subsections, we describe the off-line processing steps for data reduction, calibration and RFI rejection (flagging of channels affected by interference). RFI can be of a range of strengths, either narrow band or broad band, and their temporal variations can differ greatly from being transient to persistent over the period of observing. While some RFI are clearly visible in a single spectral record, some may be weak and only detectable after averaging spectra over time and frequency to reduce noise. We follow a hierarchical approach to detect and reject data corrupted by RFI, targeting the relatively stronger RFI in the pre-processing stage and progressively aim to reject weaker lines in the post-calibration processing steps.

\subsection{Pre-Processing}

The first processing step performs a median filter \cite{10.2307/2285666} separately on cross-correlation and auto-correlation spectra of each 16-record frame, using a moving spectral window of width $2~\rm MHz$ spanning over 17 independent frequency channels.  At this pre-processing stage, a threshold of $2\sigma$---twice the standard deviation---is adopted. This removes strong RFI from the data that stand out in the (1/16)-s integration spectra. 

Processing each 1-s frame separately, the unflagged spectral points at each frequency are averaged across the 16 records in the frame, separately for the cross-correlation as well as the two auto-correlation spectra.  A maximum of 16 unflagged points are averaged at each frequency channel, and if the number of points available for averaging is less than 4 at any frequency, we flag that frequency channel in the averaged spectrum corresponding to that time frame. Corresponding to each of these time frames, we also compute and record the standard deviation, $\sigma(\nu)$, at each frequency channel by computing the standard deviation from the unflagged points for that frequency channel, and also record the effective integration time for each averaged spectral measurement.

\subsection{Calibration}

At the end of the pre-processing, for each system state, we have three averaged spectra, namely one cross-correlation and two auto-correlations of the signals in the two arms of the receiver. We follow the method described in Sec.~\ref{sec:BP_cal} to calibrate the bandpass. We use the value of  $T_{\rm STEP}$ as derived in Sec.~\ref{sec:abs_cal} for absolute calibration. We perform complex operations on the cross spectra, yielding a complex calibrated spectrum in which the sky is expected wholly in the real component.

The calibrated spectra are derived from
\begin{align}
T_{\rm SPEC}(\nu) & = \frac{ T_{\rm OFF}(\nu)}{(T_{\rm ON}(\nu) - T_{\rm OFF}(\nu))} T_{\rm STEP} \nonumber \\
& =  \frac{ T_{\rm OFF}(\nu)}{T_{\rm TEMP}(\nu)} T_{\rm STEP}.
\end{align}
For each frequency channel, we also have associated estimates for RMS uncertainty $\sigma$ for $T_{\rm OFF}$ and $T_{\rm ON}$. We propagate these by computing the resulting uncertainty in $T_{\rm SPEC}$ using the following expression:
\begin{equation}
\sigma_{\rm SPEC} ={|T_{\rm SPEC}|}\sqrt{ \left( \ddfrac{\sigma_{\rm OFF}}{|\rm T_{OFF}|} \right)^2 
                                + \left( \ddfrac{\sigma_{\rm TEMP}}{|\rm T_{TEMP}|} \right)^2 \\
                                + 2\left(\ddfrac{\sigma_{\rm OFF}}{|T_{\rm OFF}|}\right)\left(\ddfrac{\sigma_{\rm TEMP}}{|T_{\rm TEMP}|}\right) 
                                \sigma_{(T_{\rm OFF}, T_{\rm TEMP})}},
\label{eq:error_eq}
\end{equation}
where $\sigma_{\rm OFF}$ and $\sigma_{\rm TEMP}$ are the standard deviations computed from the pre-processing step and $\sigma_{(T_{\rm OFF}, T_{\rm TEMP})}$ is the covariance between the two spectra. This last term is non-zero because the noise in $ T_{\rm OFF}$ and that in $T_{\rm TEMP}$ are correlated since latter is the difference between $T_{\rm ON}$ and $T_{\rm OFF}$. Larger is the calibration signal, lesser this covariance and hence smaller will be the relative importance of the third term under the square root in the above expression.

\subsection{RFI detection/rejection post calibration}

We now discuss the methods developed, and their underlying rationale, in flagging RFI on the calibrated spectra. There are various algorithms in the literature to detect outliers in a Gaussian noise-like signal \cite{2012PhDT........23O}. We choose median filtering as the preferred approach, with a threshold of $3\sigma$ for classifying any point as RFI in the post-calibration rejection of RFI, $\sigma$ being the standard deviation in the data. 

\subsubsection{RFI detection in 1D individual time frames}
\label{subsubsec:s2flag}

The first and critical step towards the detection of RFI in any spectrum is the modeling and subtraction of the best estimate for the true spectral shape, so that outliers may be recognized and rejected without introducing any systematic biases.  We then use median filtering as an outlier rejection algorithm on the residuals, which rejects high amplitude excursions on both positive and negative sides with equal probabilities.The concern is that if a strong RFI is present that locally biases the estimate of the true spectral power, median filtering of deviations might result in biased outcomes.   If RFI results in a positive bias in the estimate of the true spectral level locally, the high amplitude noise points that are positive will be preferentially flagged in that spectral region. This would result in systematic local biases in the spectra when averaged after such asymmetric clipping. Thus in the process of estimating for the true spectral shape, the algorithm design is required to ensure that any bias introduced would be at sub mK levels. 

To make an estimate for the true spectral shape---which we call the baseline---we first divide the frequency band of $40-240~\rm MHz$ into two sub-bands. We fit each sub-band with a 12-th order polynomial. The sub-band and the order of polynomial is high enough to represent foregrounds and systematics, whose expected shapes we have estimates of from modeling of the system, and 21-cm signal, for which we have predictions in the literature \cite{2016arXiv160902312C}, to $\rm mK$ accuracy.  

We denote the fit by $y_{\rm fit}$ and the data by $y_{\rm data}$.  There are two norms we adopt for optimization of the fit to the data: the $L_1$ norm or Least Absolute Deviations minimises $|y_{\rm data} - y_{\rm fit}|$, where as the $L_2$ norm or Ordinary Least Squares minimises $|y_{\rm data} - y_{\rm fit}|^2$ .  The $L_2$ norm is the best linear unbiased estimator of the coefficients in a fit \cite{Chipman2011}.  In the first pass of the RFI detection we adopt the $L_1$ norm since it is less sensitive to outliers as compared to the $L_2$ norm \cite{RePEc:eee:ejores:v:73:y:1994:i:1:p:70-75}. The residuals, obtained as the difference between data and fit, are tested for outliers using median filtering. We repeat the process after rejecting the RFI detected in the first step, again using $L_1$ fit, to improve the estimate of the true baseline and also progressively improve upon RFI rejection.  
\begin{figure}[htbp]
\begin{center}
\includegraphics[scale=0.4]{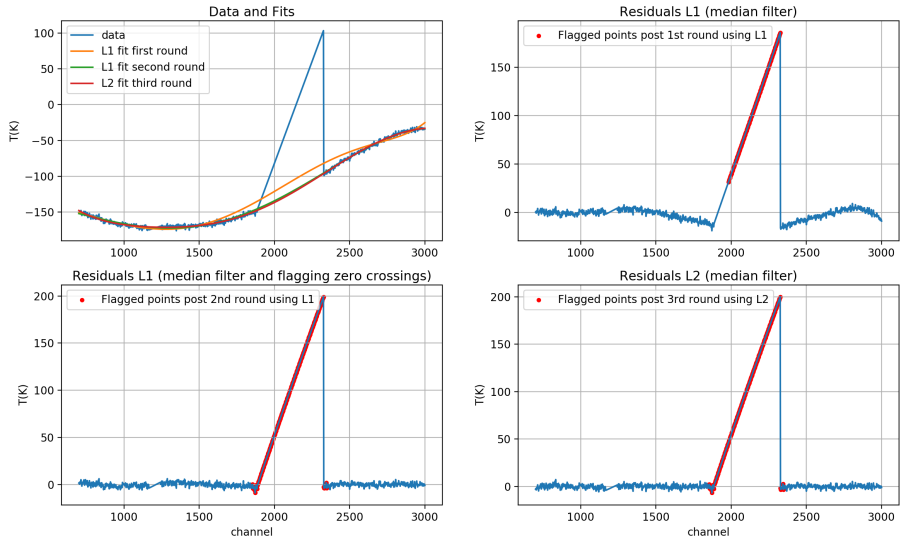}
\caption{Demonstration of the RFI rejection discussed in Sec.~\ref{subsubsec:s2flag}. The top left panel shows the mock spectrum that has an added  artefact representing a block of RFI with linearly varying strength.  The RFI detection is done in a sequential manner, as discussed in the text, and this panel shows the baseline fits at the different stages overlaid on the mock spectrum; obviously the bias in the baseline fit reduces progressively in successive stages. The upper right panel shows the residual after the first fit; the data detected as RFI at each stage is shown in red.  There is a clear structure in the residuals due to the bias in the baseline fit at this first stage.  The bias is substantially reduced when the fit is revised after rejection of RFI based in the first iteration; this is shown in bottom left panel. A median filter is applied on the residuals and in the final third iteration an $L_2$ norm based fit for a baseline is performed; the bottom right panel shows the resulting residuals. At this last stage, the entire triangular artefact is seen to be rejected.}
\label{fig:flag}    
\end{center}   
\end{figure}

RFI often appears in clusters and there is often relatively weaker RFI close to stronger RFI.  If RFI is strongly clustered, the bias in the baseline fit can be severe, and such circumstances require a different method.  To illustrate this case, consider a particularly adverse case where RFI is low in strength at one edge of an RFI cluster and progressively increases in strength towards the other end of the RFI cluster. In such a scenario, even after the two RFI rejection iterations using $L_1$ minimizations to fit for baselines, the low level RFI lying at the \textit{wings} of the cluster might still survive.  This is shown in Fig.~\ref{fig:flag}. In order to detect such low-lying RFI at the edge of an RFI cluster, we have adopted an additional data rejection step in the difference data: on each side of any rejected channel we also reject all the points along the frequency spectrum till two zero crossings of the spectral values are encountered.  This additional rejection step does inevitably cause loss of good spectral data; however, it does succeed in rejecting low levels of RFI in channels close to relatively stronger RFI.  

Following two such iterations of RFI rejection based on fitting to baselines using the $L_1$ norm, we finally perform RFI rejection using the $L_2$ minimization for estimation of the baseline followed by median filtering of residuals. Finally, as a test of the total quality of each spectrum, we compute the variance for each of the difference spectra and reject all those spectra that are outliers in their variance.  This detection of poor quality spectra is also done via a median filtering of the variance estimates.

We have carried out simulations with mock data which demonstrate that for the adopted threshold of $3\sigma$, if the offset in baselines as a result of RFI is within $20~\rm mK$ at the final stage, the bias after outlier detection will be $\sim 1~\rm mK$.  The order of the fitting polynomial and the three step process have been chosen to satisfy this tolerance.

We find that for data acquired with SARAS 2 in reasonably radio quiet sites in Ladakh in the Himalayas and in sites in South India, this process successfully rejects almost all of the obvious isolated RFI in the spectrum. 

\subsubsection{Rejection of data in 2D Time-Frequency domain}

Following the detection of RFI in the 1D individual spectra separately and sequentially, we next move to 2D time-frequency domain to detect lower levels of RFI. The strength of RFI might be lower than the median filtering threshold used on the 1D spectra, but may be detected with that confidence when the data is averaged in 2D time-frequency space. We follow a ``matched filter approach" for this. Since RFI might be spread over a time-frequency region, we progressively average the data over this 2D domain to detect lower levels of RFI as they cross the $3\sigma$ median filter threshold when the averaging enhances the amplitude of the RFI relative to the noise.

We begin once again with subtracting a baseline from each spectrum, using a fit that is an estimate of the foreground, systematics and any 21-cm signal. We divide the total spectrum in three overlapping sub-bands and separately fit each segment with 10-th order polynomials. We construct a single residual spectrum using the three residual segments, avoiding using the edges of each segment where the fits sometimes diverge from the data. 

This is done for all the spectra in the dataset yielding a 2D image of residuals over the entire time-frequency domain of the dataset. The next step of the processing is a median filtering of the entire dataset in 2D time-frequency to detect outliers. We then average the data both in time and frequency using moving windows of different widths, which progressively grow with each iteration, and perform a two-dimensional median filtering following each averaging. The maximum averaging window length currently used is $1~\rm MHz$ in frequency, assuming that CD/EoR signal has greater width.

In the 2D time-frequency domain detection of RFI, we avoid having a uniform threshold in temperature units for RFI detection using median filters, since different data points have different associated uncertainties.  This is because in the pre-processing, as well as successive iterations of RFI rejection described above, time-frequency data points are rejected and then the data is averaged and, therefore, different time-frequency data points have different effective integration times.  For every point, we examine its absolute value against its own uncertainty $\sigma$ and if the absolute value is larger than $3\sigma$, we reject the point as RFI. 

We also examine the integrated powers in each of the spectra using the corresponding polynomial fits, and reject spectra that have integrated powers that are $3\sigma$ outliers.  Such outliers result from wideband RFI, like lightning, that raise the overall power in the spectrum. 

\section{Performance Measures of SARAS 2}
\label{sec:per_meas}

Performance tests have been conducted in the laboratory to examine for spurious in the SARAS 2 receiver system and to evaluate whether the modeling of the system performance as described above (Eq.~\ref{eq:meas}) is accurate at the mK level.  We replace the antenna with standard precision reference loads or terminations with different reflection coefficients, $\Gamma_c$, acquire measurement data and model to examine for unaccounted spectral structure. 

We use three types of terminations with a range of complexity in their $\Gamma_c$:
\begin{itemize}
\item Precision $50~\mathrm{\Omega}$ termination: This is the most ideal case where $\Gamma_c$ is close to 0. Thus we have minimum reflections resulting in minimum additives arising from multipath propagation of receiver noise, reference noise and signal from the termination. 
\item Precision Open and Short loads: Open and short terminations are completely mismatched with the receiver, with $\Gamma_c$ of 1 and $-1$ respectively. All internal reflections of signals from receiver noise and reference are maximized, and in this case there is almost no signal from the termination itself.
\item Resistor-Inductor-Capacitor based network (RLC): To have a frequency behavior in $\Gamma_c$ similar to that of the antenna, we choose values of the resistor, capacitor and inductor so that the network resonates at $260~\rm MHz$, same as that of the SARAS 2 antenna, and the shape of $\Gamma_c$ is similar to that of the antenna. Thus all signals that reflect off this termination and suffer multipath propagation appear in the measurement data with systematic shapes that have the imprint of the frequency dependence of $\Gamma_c$.  
\end{itemize}

With each of these terminations in turn we acquired data for 10 hours in the laboratory and processed the measurement sets using the algorithms discussed in Section~\ref{sec:software}. The final set of spectra, after processing with the RFI rejection algorithms, were averaged in time to derive a single spectrum. We discuss below the modeling of these data, and the method of examining the residuals for the presence of spurious. The RMS noise in the residuals of the spectrum, after data modeling and without any spectral averaging, is in the range 15--20~mK.  

\subsection{Examining measurement data for spurious}
\label{subsec:check}

We analyze the residuals seeking to detect two forms of spurious: sinusoids and Gaussian shaped structures. Sinusoidal spurious are spread out in the spectral domain but appear as spikes in its Fourier domain while the Gaussian shaped spurious have a compact base in both the spectral and in its Fourier domain.

\subsubsection{Sinusoidal spurious}

Any sinusoid in the residual spectra would stand out as a spike in its Fourier domain. Thus, to detect the presence of spurious sinusoids, we perform Fourier transformation of the residuals to get a spectrum of Fourier amplitudes at different Fourier modes. These amplitudes of the Fourier transform, where the input is zero mean, follows a Rayleigh distribution \cite[Chapter 6]{Papoulis1981}. Thus, if any sinusoidal spurious exists in the residual that is detectable given the measurement noise, we would expect an outlier in the Rayleigh distributed amplitudes. 

We compute the cumulative distribution function for the amplitudes of the Fourier modes and inspect if the fraction of amplitudes above 2, 3 and 4 $\sigma$ are within the expectations for a Rayleigh distribution, assuming that the residuals are Gaussian random. Further, since the real and imaginary components of the Fourier transform are expected to have Gaussian distributions if the spectra are Gaussian random noise, the 2D distribution of real versus imaginary of the components in the Fourier transform would be expected to have a symmetric distribution. To quantify this, we test for the uniform distribution of phase of the Fourier transform using Chi-Square test \cite{9780471557791}. Any significant deviation from uniform distribution would imply the presence of coherent structure in the residuals. This is a second test for departure from Gaussianity in the Fourier domain. 

\subsubsection{Gaussian spurious}

We adopt a matched filtering approach to examine if the residuals contain Gaussian shaped structures. Gaussian functions, with a range of widths $\sigma$ are centered at a range of frequencies $\nu_0$ within the band.  The $1\sigma$ width is iteratively varied from $1-20~\rm MHz$ in different trials. We convolve the residuals with these Gaussian windows of various widths and positions.  At any location and for any width, if the summation over the product of the Gaussian window with the residual significantly exceeds the expectation from convolution of same window with a mock data that is Gaussian random noise, we may infer the presence of a Gaussian structure of width $\sigma$ at the frequency $\nu_0$.

\subsection{Modeling internal systematics}

We use the model based on the analysis of signal propagation in the SARAS 2 system, as given by the measurement equation in Eq.~\ref{eq:meas}, to fit to the data. As discussed in Sec.~\ref{sec:imp_mis}, we include higher order reflections that are expected to result in structure above mK in the model. Since contributions from orders higher than three are at sub-mK levels, we restrict to third order reflections ($\{l, m, n\} = 3$ in Eq.~\ref{eq:meas}). Further, because the SARAS 2 antenna efficiency is poor below $100~\rm MHz$, we restrict our analysis to the band $110-200~\rm MHz$.

The model fit to the calibrated measurement data, using the measurement equation, was done using the Nelder-Mead optimization \cite{doi:10.1093/comjnl/7.4.308}. Further, to avoid solutions that are unphysical, we appropriately constrain the parameters to be within expected ranges and also use multiple iterations of the Basinhopping algorithm \cite{doi:10.1021/jp970984n} to get parameters that are meaningful and acceptable given the hardware configuration and measured characteristics of the system components. 

We finally compute the difference between the spectrum and the fitted model to get the residuals. We then carry out the tests discussed above in Sec.~\ref{subsec:check} for examining for spurious sinusoidal or Gaussian shaped structures in the residuals.   

\subsubsection{Results from the $50~\mathrm{\Omega}$ termination data}
\label{subsec:50OHM}

\begin{figure}[htbp]
\begin{center}
\includegraphics[scale=0.4]{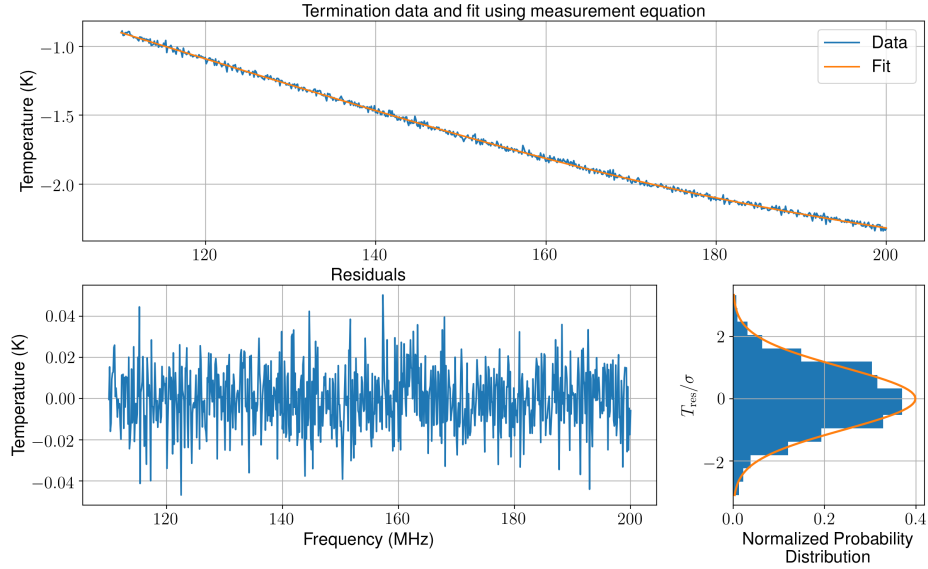}
\caption{The top panel shows the measurement data for the case of termination of the receiver with a $50~\mathrm{\Omega}$ load. The bottom left panel is the residual after subtracting the best fit model. The bottom right panel shows the distribution of the noise overlaid with a Gaussian distribution of the same mean and standard deviation.}
\label{fig:50_ohm_spec}    
\end{center}   
\end{figure}

The measured spectra, along with the residuals, for the case of modeling of measurement data from $50~\mathrm{\Omega}$ termination are shown in Fig~\ref{fig:50_ohm_spec}.  After averaging the data over the entire 10 hours, the RMS noise in the residual is $15~\rm mK$.

\begin{figure}[htbp]
\begin{center}
\includegraphics[scale=0.4]{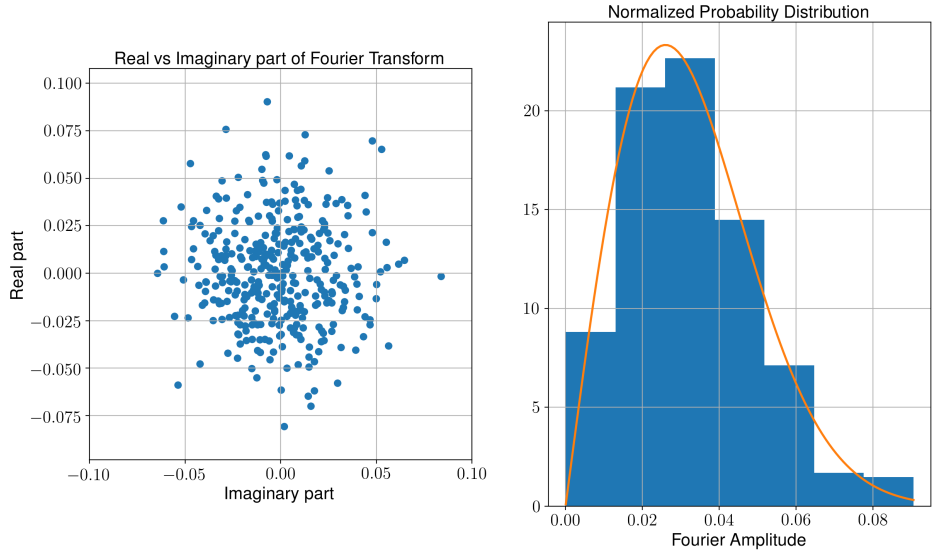}
\caption{The left panel shows real versus imaginary components of the Fourier transform values. The right panel shows the distribution of amplitudes of the Fourier transform overlaid with a Rayleigh distribution of amplitudes corresponding to the same $\sigma$. The values of real and imaginary parts are in units of $\sigma$. Both these plots are consistent with a residual containing Gaussian noise and there is no evidence for spurious.}
\label{fig:ray_50}    
\end{center}   
\end{figure}

We further test for the presence of underlying sinusoids and Gaussian structures with the methods discussed above in Sec.~\ref{subsec:check}.  We do not find any sinusoidal structures in the data down to a sensitivity of $~\sim 1~\rm mK$. This level is considerably lower than the reported RMS noise of $15~\rm mK$ since in the Fourier domain, sensitivity to any Fourier mode is enhanced by a factor that is of order the square root of half the
number of independent channels. We also show the real and imaginary parts of the Fourier transform and the distribution of these Fourier amplitudes in Fig.~\ref{fig:ray_50}. The sensitivity of the test for Gaussian structures varies with the width of the Gaussians for which the test is done.  The upper limits on amplitudes of Gaussian-type spurious range from $1-10~\rm mK$ for widths of $25~\rm MHz$ to $1~\rm MHz$ respectively. At the reported sensitivities, all the results are consistent with the measurement data being Gaussian random noise.

\subsubsection{Results from the Open/Short termination data}
\label{subsec:OPEN}

Whereas in the case of a termination that is an electrical short we expect the reflected voltage to be phase shifted by $\ang{180}$, there is no phase change on reflection from an electrical open termination. The two spectra, obtained using open and short terminations, are expected to be similar except for  this $\ang{180}$ phase shift in the reflected components. Therefore, we show the results of modeling and analysis for systematics only for the case of the open termination. 

\begin{figure}[htbp]
\begin{center}
\includegraphics[scale=0.4]{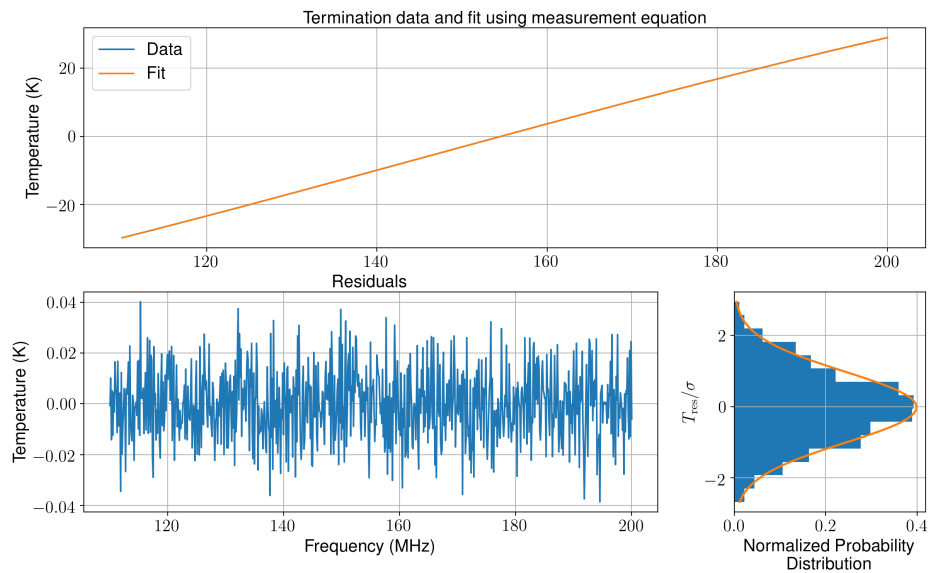}
\caption{The top panel shows the spectrum for the case of an open termination, after removal of the common additive component that is in measurement data acquired in both the open and short terminations. The bottom left panel is the residual after removing the best fit model. The bottom right panel shows the histogram of noise amplitudes overlaid with a Gaussian of same mean and standard deviation.}
\label{fig:open_spec}    
\end{center}   
\end{figure}

For both the open and short terminations, there is no source of signals at the terminations. The spectrum contains multi-order reflections from only the  reference and receiver noise $T_{\rm ref}$. We model the data using the measurement equation Eq.~\ref{eq:meas}, setting $T_a = 0$. The residuals, after subtracting the best-fit model, are shown in Fig.~\ref{fig:open_spec}.  The residuals appear consistent with Gaussian random noise, with an RMS of $\sim 15~\rm mK$. 

\begin{figure}[htbp]
\begin{center}
\includegraphics[scale=0.4]{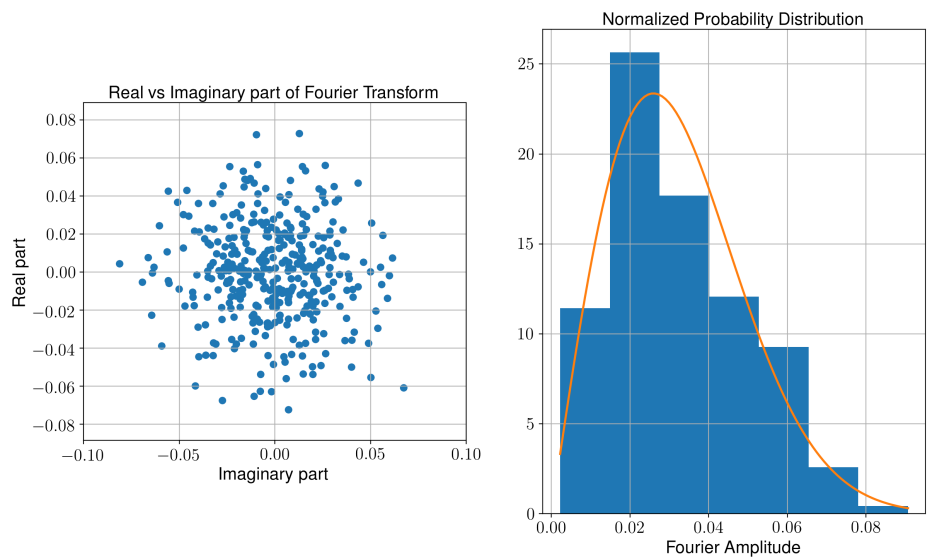}
\caption{The left panel shows the 2D distribution of real versus imaginary components of the values of the Fourier transform for residuals of the data obtained in the case of the open termination. The right panel shows the distribution of amplitudes of the Fourier transform overlaid with a Rayleigh distribution function assuming the same $\sigma$. The values of real and imaginary parts are in units of $\sigma$.}
\label{fig:open_Fourier}    
\end{center}   
\end{figure}

Similar to tests for the $50~\mathrm{\Omega}$ termination, we carried out tests for presence of sinusoids and Gaussian structures in the data for the open termination. The real and imaginary part of the Fourier Transform and the distribution of Fourier amplitudes are shown in Fig.~\ref{fig:open_Fourier}. We do not see any evidence for spurious sinusoids and Gaussian structures at the same sensitivity levels as reported in Sec.~\ref{subsec:50OHM} for the case of the $50~\mathrm{\Omega}$ termination.

\subsubsection{Results from the RLC termination data}
\label{subsec:RLC}

\begin{figure}[htbp]
\begin{center}
\includegraphics[scale=0.4]{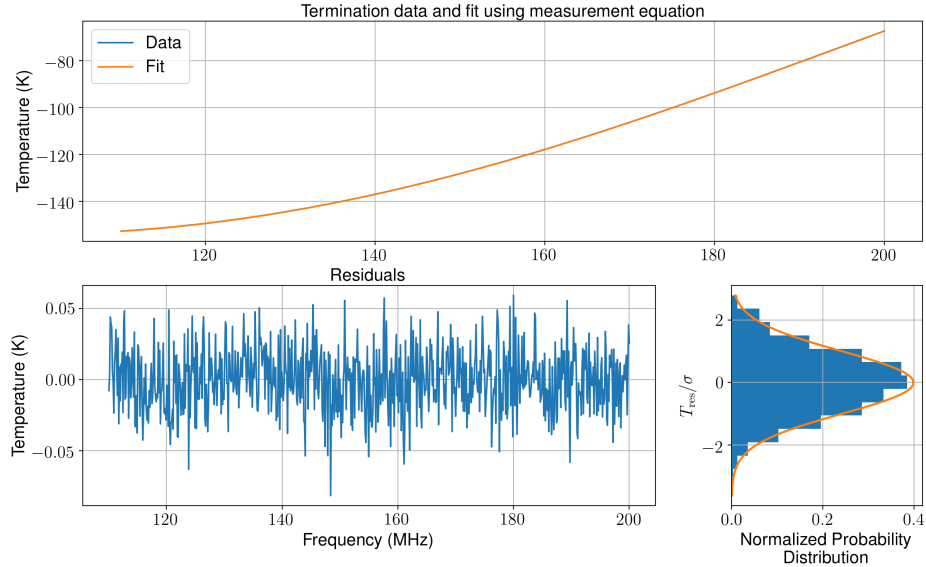}
\caption{The top panel shows the calibrated measurement data in the case of the RLC termination. The bottom left panel is the residual left after subtracting the best fit model based on the measurement equation for this termination. The bottom right panel shows the histogram of residual amplitudes  overlaid with a Gaussian with same mean and standard deviation.}
\label{fig:RLC_spectral}    
\end{center}   
\end{figure}

\begin{figure}[htbp]
\begin{center}
\includegraphics[scale=0.4]{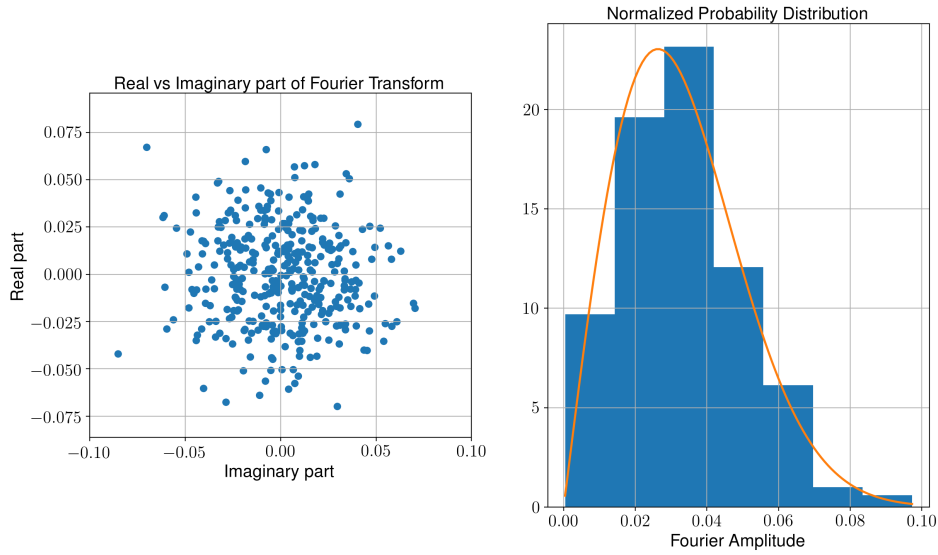}
\caption{The left panel shows 2D plot of the real versus imaginary components of the Fourier transform values for RLC termination residuals. The right panel shows the distribution of amplitudes of the Fourier transform values overlaid with Rayleigh distribution form computed for the same $\sigma$. The values of real and imaginary parts are in units of $\sigma$.}
\label{fig:RLC_Fourier}    
\end{center}   
\end{figure}

 To model the measurement data in the case of the RLC termination, we use the complete measurement equation (Eq.~\ref{eq:meas}) taking into account the spectral shape of $\Gamma_c$, coupling of $T_a$ into the system as well as multi-order reflections due to the receiver noise and also a $T_a$ equivalent from the resistance in the RLC network. We show the residuals to the model fit and its Fourier components in Fig.~\ref{fig:RLC_spectral} and \ref{fig:RLC_Fourier} respectively. The RMS noise of the residuals, after removing the best fit model from the data, is $20~\rm mK$. This is higher than $15~\rm mK$ RMS noise obtained in the other terminations. This is primarily due to more RFI flagging in case of RLC termination compared to better electromagnetically shielded $50~\mathrm{\Omega}$ and Open/Short terminations.

From the Fourier analysis, we do not find any outliers that may be evidence for sinusoids in the residual data above the measurement noise, which is $\sim 2~\rm mK$. Similarly, with the matched filter tests using different Gaussian functions, we conclude that there are no Gaussian artefacts at a level of $2-15~\rm mK$ for widths in the range $25~\rm MHz$ to $1~\rm MHz$ respectively.

\subsection{Modeling using maximally smooth functions}

Modeling using the measurement equation is challenging when the number of parameters necessary to describe the data increases, and as the complexity of $\Gamma_c$ increases. This is the case when an RLC termination replaces the antenna, with a reflection coefficient that varies across frequency, and would be the case when the antenna is connected to the system and the antenna temperature includes sky and ground radiation.  The large number of parameters in the modeling, if left free and without being determined by field or laboratory measurements, would give the model considerable freedom. There may also be degeneracy between parameters describing the model for the system and foregrounds, and confusion arising from degeneracy between parameters describing the system, foreground and 21-cm global CD/EoR signal.  This results in increased uncertainty in the derived parameters for the CD/EoR signal.  

A better approach may be to use a model description for the system, and perhaps foregrounds as well, that is less likely to subsume a substantial part of the 21 cm signal. We may thus try to approximate the calibrated measurement data with a maximally smooth function \cite{2015ApJ...810....3S}, or a variant of that which allows minimum freedom to fit out complex cosmological signals while having the necessary freedom to fit out the systematics and foregrounds. The motivation of modeling the data using a maximally smooth function is to have a limited freedom in the model such that it causes minimum loss of 21 cm signal, preserving its higher order structures, while being able to model the foreground with mK accuracy \cite{2017ApJ...840...33S}.  

\begin{figure}[htbp]
\begin{center}
\includegraphics[scale=0.4]{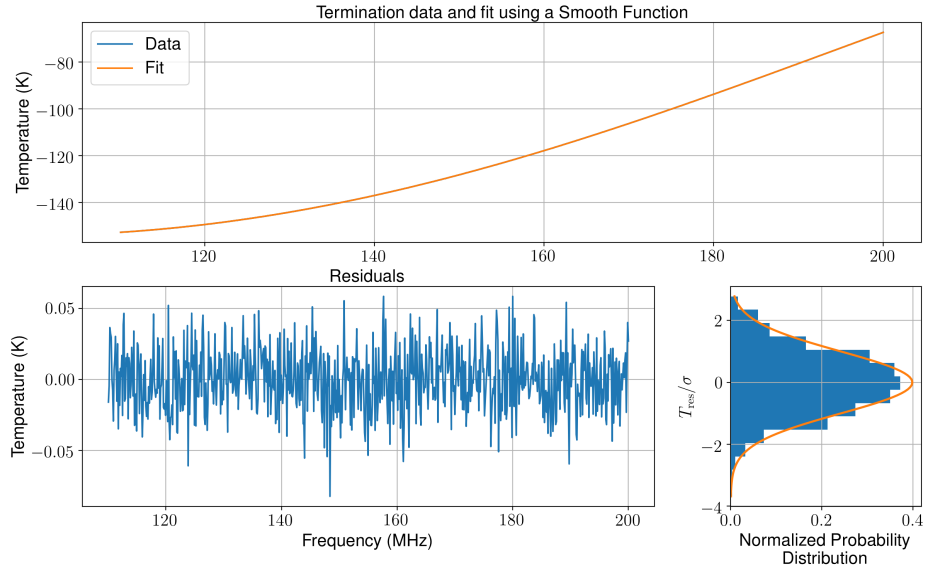}
\caption{The top panel shows the result of a maximally smooth function fit to the RLC termination data. The bottom left panel shows the corresponding residuals while the bottom right panel shows the distribution of the residuals with an overlaid Gaussian of same mean and standard deviation. Datasets with comparatively lesser complex shapes, e.g. when the termination is $50~\mathrm{\Omega}$ or Open/Short, undoubtedly reach thermal noise levels on fitting with a maximally smooth function.}
\label{fig:MS_RLC}    
\end{center}   
\end{figure}

Following this alternate approach, we fit each dataset corresponding to $50~\mathrm{\Omega}$, Open/Short and RLC terminations with maximally  smooth functions. Since there are higher order reflections of receiver and antenna signals that contribute above a mK, we allow for a maximum of one inflection in the band. With this approach, the residuals from different terminations reach the same noise levels as the residuals resulting from the fit to the data using the measurement equation in Sec. ~\ref{subsec:50OHM}, \ref{subsec:OPEN} and \ref{subsec:RLC}. We show here, as an example in Fig.~\ref{fig:MS_RLC}, the maximally smooth function fit and residuals for the data obtained with an RLC termination at the antenna terminal.

We also obtain models for the measurement data for different terminations via best fits of the measurement equation to the data. These data models capture the overall complexity of the spectrum. We have tested their smoothness by fitting these models with maximally smooth functions. The maximally smooth function is found to be able to approximate these models to $2 ~\rm mK$ level, which is a smoothness of 1 part in $10^5$ considering that the actual spectral shape may be a few hundreds of Kelvin. Thus, the tests suggest that the class of 21 cm signals that have complex variations over the band, and have antenna temperatures more than this confusion limit of $2~\rm mK$, can potentially be detected by the system.

\section{Comparison with other radiometers}

There are other ongoing experiments for detection of the global 21 cm signal from CD and EoR. In so far as we know, there are four other experiments currently underway and we list these below. 

\begin {enumerate}
\item Experiment to Detect the Global EoR Signature (EDGES) \cite{2017ApJ...835...49M}
\item Broadband Instrument for Global HydrOgen ReioNisation Signal (BIGHORNS) \cite{2015PASA...32....4S}
\item Sonda Cosmol\'ogica de las Islas para la Detecci\'on de Hidr\'ogeno Neutro (SCI-HI) \cite{2014ApJ...782L...9V}
\item Large Aperture Experiment to Detect the Dark Ages (LEDA) \cite{2017arXiv170909313P}
\end {enumerate}

We compare below the designs of these radiometers with SARAS 2; Table~\ref{tab:2} provides a summary.

SARAS 2 uses an electrically short spherical monopole antenna, targeting the signal spanning from $40-200~\rm MHz$ encompassing CD and EoR. This is in contrast with other experiments which either have different antennas and receiver systems to cover the whole band (EDGES) or target a specific subset of the band (BIGHORNS, SCI-HI and LEDA). The SARAS~2 monopole antenna ensures achromaticity of the beam over its entire wide band; however, suffers from poor efficiency at longer wavelengths. The other experiments, by employing either electrically large antennas or antennas with dimensions matched with operating frequencies, suffer from varying amounts of beam-chromaticity while maintaining good efficiency over their respective bands. 

SARAS 2 antenna design also achieves smoothness of antenna reflection coefficient to 1 part in $10^4$, which is desirable in controlling the nature of internal systematics as well as in maintaining a smooth transformation of intrinsic sky signal to the measured spectrum. In other experiments, BIGHORNS antenna has frequency structure at $10^{-1}$ level \cite{2015PASA...32....4S}. EDGES employs a separate circuitry in the field to measure the reflection coefficient and uses that in the data modeling. LEDA, on the other hand, plans to cross-correlate its radiometer with existing Long Wavelength Array (LWA) antennas and use the interferometer visibilities to solve for the antenna characteristics. Further, the SARAS 2 antenna is the only one that is a monopole and it is the only radiometer that does not have a balun, presence of which can have frequency dependent losses that are difficult to calibrate. 

Comparing receivers, other experiments essentially measure the autocorrelation of a single RF chain, and the single antenna singal is carried over to their respective digital receivers over RF cables. Further, bandpass calibration is performed by toggling between the antenna and calibration loads or noise source (Dicke switching). SARAS 2 differs from these schemes in that it employs a cross-correlation spectrometer where the signal is split into two paths immediately after it enters the receiver and the signals in the two RF paths are cross-correlated. Further, instead of toggling between load and antenna, that changes the nature of systematics in each switch position, SARAS 2 utilizes a cross-over switch where the antenna and noise source are connected all the time and hence the nature of systematics remains the same in all switch positions. The combination of cross-correlation along with the devised architecture of the receiver results in phase switching  which cancels out the spurious additive in the process of calibration.  For the transmission of RF signal from the receiver output at the antenna to the filters placed $100~\rm m$ away, SARAS 2 employs optical fibers instead of RF cables. This prevents the undesirable coupling of noise from the filters and amplifiers, which are in the signal processing unit, from one receiver arm to the other via internal reflection at the antenna; such internal reflections can result in the system noise manifesting as additive short period ripples in the measured spectrum.

\label{compare}
\begin{table}
\caption{Comparison of system designs in different experiments}
\label{tab:2}       
\begin{tabular}{|p{2cm}|p{2cm}|p{2cm}|p{1cm}|p{4cm}|p{2cm}|}
\hline\noalign{\smallskip}
Experiment & Frequency Range & Antenna & Presence of Balun & Calibration Scheme & Type of Spectrometer\\
\noalign{\smallskip}\hline\noalign{\smallskip}
SARAS 2 & $40-200~\rm MHz$ & Spherical Monopole antenna& No & Noise source coupled into system via power combiner without Dicke switch & Cross-correlation\\ [1cm]
EDGES  & $100-200~\rm MHz$ (High-Band) $50-100~\rm MHz$ (Low-Band)  & Blade Antenna & Yes & Switching between antenna and noise source & Auto-correlation\\[1.2cm]
BIGHORNS & $70 - 200~\rm MHz$ & Conical log-spiral antenna &Yes & Switching between antenna and reference load & Auto-correlation\\[0.5cm]
SCI-HI & $40 - 130 ~\rm MHz$ & HIbiscus antenna & Yes & Switching between antenna, $50~\mathrm{\Omega}, 100~\mathrm{\Omega}$ and short termination & Auto-correlation\\ [0.8cm]
LEDA & $40 - 85 ~\rm MHz$&Dual-polarized dipole antenna & Yes & Switching between antenna and noise source combined with cross-correlation from other antennas for antenna gain and beam estimation & Auto-correlation\\[1cm]
\noalign{\smallskip}\hline
\end{tabular}
\end{table}

\section{Summary}

We have developed a wideband precision spectral radiometer, SARAS 2, towards detection of 21 cm global signal from Cosmic Dawn and the Epoch of Reionization in the frequency range $40-200~\rm MHz$. For each sub-system, as well as for various data processing strategies, we have discussed the favourable features that would aid in the detection. Using these criteria, we have evolved the radiometer design to have characteristics conducive to the experiment.

The electromagnetic sensor is a spherical monopole antenna, with a frequency independent beam along with a spectrally smooth reflection and radiation efficiencies. The properties of the antenna have been characterized by simulations and field measurements. In the process, we have also developed a novel way of measuring the total antenna efficiency using GMOSS and acquired sky data. The analog receiver has been designed such that the system can be calibrated without Dicke switching, along with a mechanism to cancel the spurious internal additives through signal splitting and cross-correlation. The receiver is connected directly to the antenna thereby minimizing the lengths in the system, which otherwise would result in high order frequency structure. The configuration has been devised to control the nature of internal systematics and keep them spectrally smooth in order to discern between foreground, systematics and the 21 cm signal. 

We have outlined the signal path in the system leading to the measurement equation, including multi-order reflections, along with the description and the rationale  of  the algorithms developed for data pre-processing, calibration and RFI rejection. This is followed by the evaluation of system performance by connecting various terminations replacing the antenna, with increasing complexities of the resulting systematics. 

We have analyzed the internal systematics by using the measurement equation to approximate the spectrum as well as by modeling it with a maximally smooth function. Using both these methods, we get data residuals with RMS noise ranging from 15--20~mK for 10~hr of integration for all terminations. The residuals, which are dominated by Gaussian noise, are then tested for the presence of sinusoids and Gaussian shaped structures. Using various tests developed, we place an upper limit of $2~\rm mK$ for sinusoidal spurious and an upper limit of 2--15~mK for Gaussian shaped structures with width in the corresponding range of 20--1~MHz. 

Thus the system has been demonstrated to be sensitive to $\rm mK$ levels without being limited by any un-modeled systematic structure, consistent with the requirements of radiometer design for precision measurements of global cosmological redshifted 21 cm from CD/EoR. The future upgrade in SARAS is on improvement of the antenna total efficiency without compromising the existing properties. 

\begin{acknowledgements}

We thank RRI Electronics Engineering Group, particularly Kasturi S., Madhavi S. and Kamini P. A., for their assistance in analog and digital receiver assembly. We also thank the Mechanical Engineering Group (RRI), led by Mohamed Ibrahim, for manufacturing the antenna along with construction of chassis and shielding cages for analog and digital receivers. Santosh Harish and Divya Jayasankar took an active role in developing real-time data acquisition software and system monitoring hardware respectively. We are grateful to the staff at the Gauribidanur Field Station led by H.A. Ashwathappa for providing excellent support in carrying out field tests and measurements.

\end{acknowledgements}

\begin{appendices}

\section{Measurement of the total efficiency of an antenna using the Global Sky Model}
\label{app:1}

We describe here a method developed to derive the total efficiency of any antenna using a night sky observation with a radiometer.  The method uses a global model for the sky brightness distribution. We have used GMOSS \cite{2017AJ....153...26S} as the model and the SARAS 2 receiver to measure the total efficiency of the SARAS 2 monopole antenna.

At any frequency, the calibrated measurement data can be decomposed into a sum of contributions from the foreground, ground and receiver systematics. Further, the data is measured with reference to a standard load, whose physical temperature over the observing time is recorded using a logger. The contributions from foreground and the reference load temperature are the only significant time-varying components in the data. Since the instrument is being calibrated every second, all the temporal variations in receiver gain are calibrated out. Though the ground temperature may vary over the observing time, this is a variation on the surface; the effective temperature of the ground emission corresponds to the temperature at an effective penetration depth. For ground contribution in the frequencies of interest, the effective penetration depth is $\sim 2.5~\rm m$ \cite[Chapter 3]{9780872599871}. At this depth, the diurnal temperature variations have been found to be negligible \cite{Florides_1annual}. Thus the contributions from instrument systematics as well as contributions from the ground are essentially time invariant and may be treated as constant additives in the spectrum. 

The measured temperature, thus, can be represented as :
\begin{equation}
T_A(\nu, t) = \eta_t(\nu) T_{\rm W}(\nu, t) + T_{\rm add}(\nu) - T_{\rm ref}(t),
\end{equation}
where $T_A$ is the measured power, $T_{\rm W}$ is the beam-weighted foreground that couples to the system through the total efficiency $\eta_t$ of the antenna, and $T_{\rm ref}$ is the reference temperature. It is to be noted that we actually measure the physical temperature of the reference load $T_{\rm p}$, which is linearly related to the actual temperature $T_{\rm ref}$. This is because there is a thermal resistance between the actual source of noise and the outer metallic body where the temperature is measured. The time-invariant component of the data consisting of the systematics and ground contributions is represented by $T_{\rm add}$.

Using GMOSS, at every frequency, we decompose the time series data at each frequency into three components:
\begin{itemize}
\item a component correlated with temporal variations in the foreground brightness,
\item a component correlated with temporal variations in the reference load temperature, and
\item a component that is constant over time.
\end{itemize}
Thus, at any frequency $\nu$ we have the following equation:
\begin{equation}
T_A(\nu, t) = \eta_t(\nu)T_{\rm W}(\nu, t) + a_1 T_{\rm p}(t) + a_2(\nu),
\end{equation}
where $T_A$ is the measured power, $T_{\rm W}$ is derived from GMOSS as a weighted average of the model $T_B$ over the sky with a weighting by the antenna beam. $T_{\rm p}$ is taken from the reference load temperature measurements. Using these, we optimize for $a_1$ using a measurement data across time and frequency that includes sufficient LST range so that the antenna temperature varies significantly.  $\eta_t$ and $a_2$ are optimized for each frequency independently.

\begin{figure}[htbp]
\begin{center}
\includegraphics[scale=0.3]{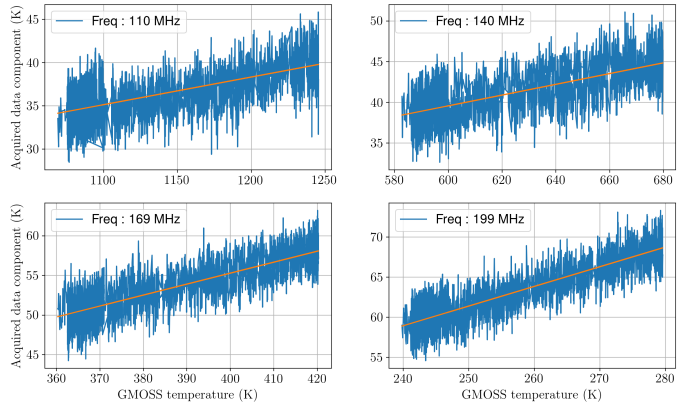}
\caption{The component of data that is correlated with GMOSS foreground predictions shown at four sample frequencies. The slope of the line at each frequency provides an estimate of the total efficiency at that frequency.}
\label{fig:fig_fore}    
\end{center}   
\end{figure}

The resulting total efficiency across the band is shown in Fig.~\ref{fig:fig_effp}. We also show the optimization fits for total efficiency at four representative frequencies in Fig.~\ref{fig:fig_fore}. This method also provides an estimate of the additives in the system, $a_2(\nu)$, which may be used as a tool to model the data.

\end{appendices}

\bibliographystyle{spmpsci}       
\bibliography{references}   

\end{document}